\documentclass[10pt,twocolumn]{IEEEtran}

\usepackage{amsmath,amsfonts,graphics,amssymb,epsfig,subfigure,color,cite,xcolor}                             
\usepackage{amsthm,textcomp}
\theoremstyle{plain}
\newtheoremstyle{mystyle}%                % Name
  {0mm}%                                  % Space above
  {0mm}%                                  % Space below
  {}%                                     % Body font
  {4mm}%                                  % Indent amount
  {\bfseries}%                            % Theorem head font
  {:}%                                    % Punctuation after theorem head                  
  { }%                                    % Space after theorem head, ' ', or \newline
  {\thmname{#1}\thmnumber{ #2}\thmnote{ (#3)}}%                                     % Theorem head spec (can be left empty, meaning `normal')
\theoremstyle{mystyle}
\usepackage{bm}
\usepackage{array}
\usepackage{multirow}
\usepackage{enumerate}
\usepackage{algorithm}
\usepackage{algpseudocode}
\usepackage{adjustbox}
\usepackage{algcompatible}
\algblockdefx{FORP}{ENDFORP}[1]%
  {\textbf{for }#1 \textbf{do in parallel}}%
  {\textbf{end}}
\algblockdefx{FOR}{ENDFOR}[1]%
  {\textbf{for }#1 \textbf{do}}%
  {\textbf{end}}
\algblockdefx{noEnd}{noEndEnd}[1]
    {#1}
    {{\Statex}}
\algblockdefx{BULLET}{EndBULLET}{$\bullet$ }{}
\algnotext{EndBULLET}
\algblockdefx{IF}{ENDIF}[1]%
  {\textbf{if }#1 \textbf{then}}%
  {\textbf{end}}
\algnewcommand\algorithmicprocedure{\textbf{function}}
\algnewcommand\FUNC{\item[\algorithmicprocedure]}%
\algnewcommand\algorithmicendprocedure{\textbf{end function}}
\algnewcommand\ENDFUNC{\item[\algorithmicendprocedure]}%
 \usepackage{setspace}
\let\Algorithm\algorithm
\renewcommand\algorithm[1][]{\Algorithm[#1]\setstretch{1.4}}
\usepackage{float}
\usepackage{makecell}
\usepackage{graphicx}
\usepackage{mathrsfs}
\usepackage[caption=false,font=normalsize,labelfont=sf,textfont=sf]{subfig}
\usepackage[right=0.625in, left=0.625in, top=0.7in, bottom=1.0in]{geometry}
\usepackage{verbatim}

\usepackage{bbm}

\floatname{algorithm}{Procedure}

\makeatletter
\newcommand{\vast}{\bBigg@{4.5}}
\newcommand{\Vast}{\bBigg@{7.5}}
\makeatother

\allowdisplaybreaks

\begin{document}
    % \title{Hierarchical Framework for Wideband Integrated Sensing and Communications with Uniform Planar Array}
    \title{Beam-Squint-Aided Hierarchical Sensing for Integrated Sensing and Communications with Uniform Planar Arrays}
    % \title{Wideband ISAC via Hierarchical Sensing Framework for UPA}
    % \title{Location-Aware Robust Beam Alignment \\ for Low-SNR Millimeter-Wave Communications}
 %    \author{\IEEEauthorblockN{Jihun Park, Yongjeong Oh, Jaewon Yun, Seonjung Kim, and Yo-Seb Jeon}
	% \IEEEauthorblockA{Department of Electrical Engineering, POSTECH, Pohang, Gyeongbuk 37673,          Republic of Korea \\
	% Email: \{jihun.park, yongjeongoh, jaewon.yun, seonjung.kim, yoseb.jeon\}@postech.ac.kr}
 %        }
    \author{Jaehong Jo, Jihun Park, Yo-Seb Jeon, \IEEEmembership{Member,~IEEE}, and H. Vincent Poor, \IEEEmembership{Life Fellow,~IEEE}
    \thanks{Jaehong Jo, Jihun Park, and Yo-Seb Jeon are with the Department of Electrical Engineering, POSTECH, Pohang, Gyeongbuk 37673, South Korea (e-mail: jaehongjo@postech.ac.kr; jihun.park@postech.ac.kr; yoseb.jeon@postech.ac.kr).}
    \thanks{H. Vincent Poor is with the Department of Electrical Engineering, Princeton University, Princeton, NJ 08544 (e-mail: poor@princeton.edu).}
    % \thanks{\textcolor{red}{This work was presented in part at the 2022 IEEE Global Communications Conference (GLOBECOM)}.}
    }
    \vspace{-2mm}	
    
    \maketitle
	% \makeatletter
 %    \def\blfootnote{\gdef\@thefnmark{}\@footnotetext}
 %    \makeatother
    %\vspace{-12mm}

    \begin{abstract} % up to 200 words
        In this paper, we propose a novel hierarchical sensing framework for wideband integrated sensing and communications with uniform planar arrays (UPAs). Leveraging the beam-squint effect inherent in wideband orthogonal frequency-division multiplexing (OFDM) systems, the proposed framework enables efficient two-dimensional angle estimation through a structured multi-stage sensing process. Specifically, the sensing procedure first searches over the elevation angle domain, followed by a dedicated search over the azimuth angle domain given the estimated elevation angles. In each stage, true-time-delay lines and phase shifters of the UPA are jointly configured to cover multiple grid points simultaneously across OFDM subcarriers. To enable accurate and efficient target localization, we formulate the angle estimation problem as a sparse signal recovery problem and develop a modified matching pursuit algorithm tailored to the hierarchical sensing architecture. Additionally, we design power allocation strategies that minimize total transmit power while meeting performance requirements for both sensing and communication. Numerical results demonstrate that the proposed framework achieves superior performance over conventional sensing methods with reduced sensing power.
    \end{abstract}
    
 
    \section{Introduction}\label{Sec:Intro}
    % Motivation for ISAC in 6G
    Emerging wireless standards aim to support diverse applications such as autonomous driving, smart manufacturing, and immersive extended reality \cite{ISAC_appl1, ISAC_appl2}. These applications demand ultra-reliable, low-latency communication, high data rates, and precise sensing capabilities for accurate perception and localization. Traditionally, wireless communications and radar sensing have operated independently, leading to inefficient spectrum utilization and considerable hardware cost. Integrated sensing and communications (ISAC) has emerged as a promising solution, unifying sensing and communication within shared available resources e.g., frequency spectrum, transmit power, and transmit time duration. Beyond resource sharing, ISAC enables communication and sensing to assist each other through real-time environmental data acquisition, facilitating dynamic network management and optimal beamforming through precise parameter estimation \cite{Mutual_assist1,mutual_assist2}. Multi-modal sensing further enhances detection and localization accuracy by integrating data from various sensor sources, making it highly effective in complex environments \cite{Multi_modal}. These advantages have established ISAC as a key usage scenario for IMT-2030/6G \cite{IMT2030_1, IMT2030_2}.
    
    % Motivation for beamforming >>
    Accurate beamforming using an antenna array plays a pivotal role in enabling ISAC systems, offering several key advantages for enhancing the performance of both communication and sensing. By generating narrow and directional beams, beamforming allows the system to precisely estimate target positions while achieving high spatial resolution. It also mitigates undesired interference by concentrating energy in the desired direction \cite{UPA1}. 
    % For both communication and sensing, accurate beamforming provides significant gain, effectively compensating for path loss. 
    Motivated by these benefits, fully digital beamforming techniques for ISAC systems have been investigated in \cite{digital1,digital2,ULA2,ULA3,ULA5}. Leveraging baseband digital signal processing, these techniques offer great flexibility in jointly optimizing beam patterns and waveform designs. However, fully digital approaches require as many RF chains as antenna elements, leading to prohibitive hardware costs and power consumption \cite{drawback_digital}.
        
    To address this challenge, hybrid beamforming techniques for ISAC systems have been developed in \cite{hybrid1, hybrid3}, aiming to reduce both hardware cost and power consumption without significantly compromising performance. An alternating optimization-based hybrid beamforming method was proposed for orthogonal frequency-division multiplexing (OFDM)-based ISAC systems in \cite{hybrid1}, while \cite{hybrid3} explored a partially connected hybrid architecture designed to minimize the Cramér$-$Rao bound (CRB) for sensing. Most of these studies have focused on uniform linear arrays (ULAs) due to their structural simplicity and analytical tractability. 
    However, the ULA structure is fundamentally limited as it lacks the capability to resolve spatial angles in terms of both azimuth and elevation. This limitation can be addressed by a uniform planar array (UPA), which introduces a critical second degree of freedom for angle estimation that enables direct two-dimensional (2D) localization from a single aperture, a crucial advantage for achieving more accurate spatial perception in complex environments.
    Motivated by these advantages, ISAC systems with a UPA have gained increasing attention in recent research \cite{UPA1,UPA2,UPA3}. For instance, \cite{UPA1} analyzed the CRB performance of a UPA-based ISAC system, while \cite{UPA3} investigated a near-field ISAC scenario with an extremely large UPA, designing beamformers to minimize the sum CRB. Despite these benefits, conventional grid-based localization using UPAs often suffers from prohibitive scanning time due to the necessity of exhaustive searches over a large 2D angular space, leading to significant sensing latency.

    The adoption of wideband signaling is another pivotal enabler. This approach unifies both sensing and communication functionalities on the same OFDM waveform, thereby enhancing sensing resolution and data transmission rates simultaneously. Wideband signals enhance both sensing resolution and data transmission rates simultaneously \cite{wideband}. However, as the signal bandwidth increases, the deviation of the beam direction across subcarriers---known as the {\em beam-squint} effect---becomes inevitable \cite{beam_squint_solve}. This effect poses a significant challenge in maintaining accurate and narrow beams across the entire bandwidth, leading to degraded beam alignment across subcarriers. To address this, traditional methods have primarily focused on mitigating the beam-squint effect through digital-domain phase correction, subcarrier-dependent-phase shifters (SD-PSs), beam broadening, or true-time-delay (TTD) lines \cite{beam_squint_solve}. While digital phase correction and SD-PSs can reduce the beam-squint effect, they do so at the cost of hardware efficiency. Beam broadening, on the other hand, sacrifices array gain.% wideband ISAC systems,

    A more promising direction is to exploit the beam-squint effect to design favorable beam patterns rather than suppressing it. Based on this idea, active beamforming techniques using TTD lines have been investigated in \cite{YOLO,beam_squint_split,use_beam_squint1,use_beam_squint2,use_beam_squint3,use_beam_squint4}, with the aim of using frequency-dispersed beams to improve the sensing beam sweep efficiency in ISAC systems. Most of these studies consider ULA architectures, with some specifically focusing on near-field scenarios \cite{use_beam_squint2, use_beam_squint3, use_beam_squint4}. For ISAC systems with the UPA, a joint design of TTD, PS, and power allocation was studied based on a deep learning approach \cite{use_beam_squint_UPA1}. Furthermore, in \cite{use_beam_squint_UPA2}, the trade-off between communication rates and sensing coverage was analyzed when exploiting the beam-squint effect in terahertz OFDM-ISAC systems. Despite these efforts, the problem of designing active beam patterns that can efficiently cover the elevation–azimuth angle domain formed by the UPA remains an open challenge. The core difficulty in this beamforming design lies in a fundamental trade-off between enlarging angular coverage by leveraging beam dispersion and increasing beamforming gain by mitigating it. Furthermore, designing a subcarrier-level power allocation strategy to ensure uniform sensing quality remains unsolved in prior studies.
    
    %\subsection{Contributions}
    To fill this research gap, this paper proposes a novel hierarchical sensing framework for UPA-based ISAC systems that simultaneously addresses the high computational complexity of conventional grid-based sensing and the beam-squint effect inherent in wideband operations by explicitly exploiting beam-squint as a sensing resource. Unlike prior works that separately adopted wide or horizontal beamforming \cite{use_beam_squint4}, \cite{beam_broad1,beam_broad2}, our framework strategically integrates them into a structured process that systematically decouples the 2D angle estimation into sequential elevation and azimuth searches. At each stage, frequency-dispersed beams are generated via the beam-squint effect to simultaneously cover $N$ distinct angles. For each search process, we develop a target detection algorithm inspired by sparse signal recovery in compressed sensing. Additionally, we optimize power allocation to ensure consistent performance across both sensing and communication. Simulation results demonstrate that the proposed framework significantly reduces sensing power through hierarchical processing, while achieving superior performance over exhaustive-based and subspace-based methods.
    
    The major contributions of this paper are summarized as follows:
    \begin{itemize}
    \item We introduce a novel hierarchical sensing framework that performs decoupled 2D angle estimation through two sequential processes consisting of (i) elevation angle sensing (EAS) and (ii) azimuth angle sensing (AAS). This hierarchical dependence enables this process without exhaustive $N \times N$ grid searches, thereby significantly reducing sensing time and complexity. In every process, the beam-squint effect is judiciously exploited by jointly controlling the TTD lines and PSs to simultaneously cover multiple elevation or azimuth grid points across the subcarriers. In contrast to sensing beamforming, communication beamforming is designed to maximize the beamforming gain at each user’s location by compensating for the beam-squint effect across all subcarriers.

    \item We develop a practical target detection strategy for the proposed hierarchical sensing framework, enabling efficient estimation of the elevation and azimuth angles of multiple targets. To achieve this, we formulate the estimation problem at each stage as a sparse signal recovery problem in compressed sensing \cite{CS_book}. In this formulation, the sensing beam gain pattern serves as the measurement vector, while an unknown phasor vector is treated as a sparse vector with unit-magnitude nonzero entries. Based on this interpretation, we adopt a greedy approach by modifying the conventional matching pursuit (MP) algorithm to solve the problem effectively \cite{MP}.
    
    \item We develop a sensing power allocation strategy aimed at minimizing the total transmit power for sensing, subject to a sensing signal-to-noise ratio (SNR) constraint. This strategy ensures consistent sensing performance across different target locations. Building on this, we also design a communication power allocation strategy that minimizes its transmit power while satisfying a communication signal-to-interference-plus-noise ratio (SINR) constraint that explicitly incorporates interference from the sensing signals. This joint optimization guarantees that communication performance requirements are met, even under the coexistence of both communication and sensing operations.

    \end{itemize}

    This work extends our previous study \cite{GC_Jo}, which was fundamentally limited by its focus on single-target sensing. In this current work, we address this limitation by developing a more general and comprehensive framework designed for the complexities of multi-target environments. Concurrently, we advance the underlying compressed sensing problem formulation to achieve sensing performance with higher resolution. We further extend our previous work by introducing power allocation strategies for sensing and communication, which was not explored in \cite{GC_Jo}. These additional contributions address the critical challenge of minimizing the total transmit power while guaranteeing a minimum sensing accuracy and a minimum communication rate.

    % {\em Organization:} The remainder of this paper is organized as follows. Section~\ref{Sec:System} describes the ISAC system considered in this work, including the channel and signal models. Section~\ref{Sec:Hierarchical} provides an overview of the proposed hierarchical sensing framework and details its beamforming design. In Section~\ref{Sec:Detection}, we develop a target detection strategy tailored to the proposed framework. Section~\ref{Sec:Power} introduces a power allocation strategy that ensures efficient operation for both sensing and communication. Simulation results are presented in Section~\ref{Sec:Simulation} to validate the proposed approach. Finally, Section~\ref{Sec:Conclusion} concludes the paper.
    
    {\em Notations:} Uppercase boldface $ \bf X$, lowercase boldface $ \bf x$, and lowercase non-bold $x$ denote matrices, vectors, and scalars, respectively. The sets of complex numbers, real numbers, and natural numbers are represented by  $\mathbb{C}$, $\mathbb{R}$, and $\mathbb{N}$, respectively. The Kronecker product is denoted by $\otimes$. For a scalar $x$, the absolute value is denoted by $|x|$, and the ceiling function is denoted by $\lceil x \rceil$ if $x \in \mathbb{R}$. For a vector $ \bf x$, its conjugate, transpose, and conjugate transpose are denoted by $\mathbf x^{*}$, $\mathbf x^{\sf T}$, and $\mathbf x^{\sf H}$, respectively. The operator ${\sf diag}(\mathbf x)$ constructs a diagonal matrix whose diagonal entries are the components of $ \bf x$. The $\ell_p$-norm is denoted by $\| \mathbf x \|_p$, where $p \geq 0$. For a matrix $\mathbf X$, the $k$-th column is denoted by $[\mathbf X]_{:,k}$ and the $(i,j)$-th entry is denoted by $[\mathbf X]_{i,j}$. The sets $\{ x(k) \}_k$ and $\{\mathbf{x}(k) \}_k$ represent collections of scalars and vectors, respectively, indexed by $k$.
    
    % $\mathbf X(:,k)$. , A set $\{ x(k) \}_k$ represents a collection of scalar values indexed by $k$. , The $i$-th component of $ \bf x$ is denoted by $ [\mathbf x]_{i}$. 
    
    \section{System Model}\label{Sec:System}

    \begin{figure}[t]
        \centering %\vspace{-3mm}s
        {\epsfig{file=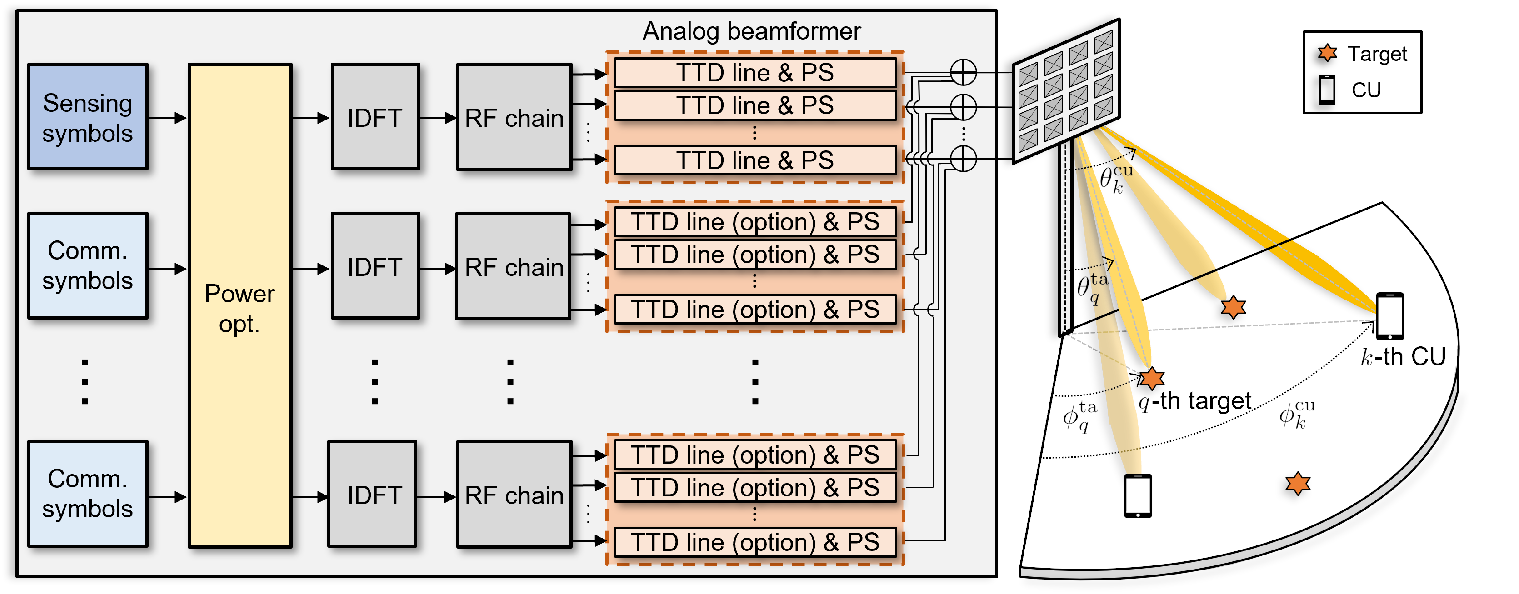,width=9cm}}%\vspace{-3mm}
        \caption{An illustration of an ISAC system where the BS equipped with UPAs communicates with $K$ single-antenna CUs and senses $Q$ passive targets.}  %\vspace{-3mm}
        \label{fig:system_model}
    \end{figure}
    
    %\subsection{Considered Scenario}
    Consider a monostatic ISAC system that uses wideband signals. A single base station (BS), equipped with two UPAs for transmission and reception, transmits information signals to $K$ single-antenna communication users (CUs) while simultaneously estimating the positions of $Q$ targets in the region-of-interest (ROI), as illustrated in Fig.~\ref{fig:system_model}. The $Q$  targets are assumed to be passive, acting solely as reflectors of sensing signals transmitted by the BS. The ROI is defined by an elevation angle range from $\theta_{\text{min}}$ to $\theta_{\text{max}}$ and an azimuth angle range from $\phi_{\text{min}}$ to $\phi_{\text{max}}$. 
    Each UPA of the BS is positioned at a height of $H$ and comprises $M = M_{\rm h} \times M_{\rm v}$ antenna elements with half-wavelength spacing, where $M_{\rm h}$ and $M_{\rm v}$ are the numbers of antenna elements in the horizontal and vertical axes, respectively. The BS is assumed to have $K+1$ RF chains, and each RF chain is connected to an analog beamformer that consists of $M$ PS and $M$ passive TTD lines\footnote{It is noteworthy that the TTD lines for the communication RF chains can be considered optional. This is because the beam-squint effect becomes negligible when the fractional bandwidth (i.e., the ratio of the bandwidth to the carrier frequency) is small. In such scenarios, the TTDs may not be essential for the communication links.}, as illustrated in Fig.~\ref{fig:system_model}. Specifically, $K$ RF chains are utilized for transmitting information signals to $K$ CUs, while a single dedicated RF chain transmits a sensing signal for target detection. Details of this setting are elaborated in Sec.~\ref{Sec:Hierarchical}.

    In this work, we consider a full-duplex scenario in which self-interference is suppressed using an advanced cancellation technique \cite{digital2}. The BS is assumed to have accurate location information for each CU, and the CUs are distributed with adequate spatial separation. Both the CUs and the targets are supposed to remain stationary throughout the entire sensing procedure \cite{digital1, user_loca1,user_loca2}. This setup enables the development of a flexible and reliable joint sensing and communication framework, which can be extended to incorporate dynamic CU locations in future scenarios.

    \subsection{Channel Model with Beam-Squint Effect}
    The BS array aperture is assumed to be much smaller than the BS-CU distance and BS-target distance, ensuring that the wavefronts can be treated as planar at the CUs and targets (i.e., far-field assumption). The system utilizes an OFDM waveform for both communication and sensing purposes. The transmission bandwidth and carrier frequency are denoted by $F$ and $f_c$, respectively, with the subcarrier frequencies ranging from $f_c$ to $f_c + F$. Given a total of $N$ subcarriers, the first subcarrier has the lowest frequency $f_1=f_c$, while the frequency of the $n$-th subcarrier is denoted by $f_n = f_c + \tilde{f}_n$, where $\tilde{f}_n = \frac{(n-1)F}{N-1}$.
    We denote $\tilde{f}$ as a frequency deviation from the carrier frequency. 
    Due to the structure of the UPA at the BS, the normalized frequency-dependent steering vector that incorporates beam-squint effect relative to the frequency $\tilde{f}$ can be expressed using horizontal and vertical steering vectors as follows \cite{UPA1}:
    \begin{align}\label{eq:vectorized_steering_vector}
        \mathbf a(\theta, \phi,\tilde{f}) =  \mathbf a_{\rm h}(\theta, \phi,\tilde{f}) \otimes \mathbf a_{\rm v}(\theta, \tilde{f}),
    \end{align} 
    where $\mathbf a_{\rm h}(\theta, \phi,\tilde{f})$ is the horizontal steering vector, given by 
    % \begin{align}\label{eq:M_h_steering_vector}
    %     \mathbf a_{\rm h}(\theta,\phi,\tilde{f}) = \frac{1}{\sqrt{M_{\rm h}}}\Big [& 1,e^{-j \pi \sin \theta \cos \phi (1+\frac{\tilde{f}}{f_c})}, \nonumber \\ & \ldots, e^{-j \pi (M_{\rm h} -1)\sin \theta \cos \phi (1+\frac{\tilde{f}}{f_c})} \Big ],
    % \end{align}
    \begin{align}\label{eq:M_h_steering_vector}
        \mathbf a_{\rm h}(\theta,\phi,\tilde{f}) = \frac{1}{\sqrt{M_{\rm h}}}\Big [ & 1,e^{-j \pi \sin \theta \cos \phi (1+\frac{\tilde{f}}{f_c})}, \nonumber \\ 
        & \ldots, e^{-j \pi (M_{\rm h} -1)\sin \theta \cos \phi (1+\frac{\tilde{f}}{f_c})} \Big ],
    \end{align}
    \begin{align}\label{eq:M_v_steering_vector}
            \mathbf a_{\rm v}(\theta,\tilde{f}) =\frac{1}{\sqrt{M_{\rm v}}}\Big [&  1,e^{-j \pi \cos \theta (1+\frac{\tilde{f}}{f_c})}, \nonumber \\ 
            & \ldots, e^{-j \pi (M_{\rm v} -1)\cos \theta (1+\frac{\tilde{f}}{f_c})} \Big ]. 
    \end{align}
    %,, determined exclusively by the azimuth angle $\phi$,  s given by 
    % ensuring that $\tilde{f}_1 = 0$ and $\tilde{f}_N = F$. 
    %By setting the positions corresponding to the frequency $\tilde{f}$ to zero, the steering vector that ignores the beam-squint effect can be derived.
    Note that the horizontal steering vector is determined by both the elevation angle $\theta$ and the azimuth angle $\phi$, while the vertical steering vector is determined exclusively by the elevation angle $\theta$.
    %Utilizing the above notations, the normalized frequency-dependent steering vector can be expressed as 
    Let $\theta_q^{\rm ta}$ and $\phi_q^{\rm ta}$ be the elevation and azimuth angles of the $q$-th target, respectively. The round-trip line-of-sight (LoS) component between the BS and the $q$-th target on subcarrier $n$ is characterized as
    \begin{align}\label{eq:LOS_component_Sensing}
        \mathbf G^{\rm los}_{n}(\theta_q^{\rm ta}, \phi_q^{\rm ta}) = \alpha_q e^{-j 4\pi l^{\rm ta}_q/\lambda} \mathbf a^{\sf H}(\theta_q^{\rm ta}, \phi_q^{\rm ta},\tilde{f}_n) \mathbf a(\theta_q^{\rm ta}, \phi_q^{\rm ta},\tilde{f}_n),
    \end{align}
    where $\alpha_q = \sqrt{\frac{\lambda^2 M^2 \sigma_{\text{RCS}}}{(4 \pi)^3 (l^{\rm ta}_{q})^4}}$ is a sensing channel attenuation gain \cite{YOLO}, $\lambda$ denotes the wavelength, $l^{\rm ta}_q = H / \cos \theta_q^{\rm ta}$ represents the distance between the BS and the $q$-th target, and $\sigma_{\text{RCS}}$ is a target radar cross section (RCS).
    To capture cluttered non-line-of-sight (NLoS) returns, we adopt a Rician fading model for the overall sensing channel on subcarrier $n$, $\mathbf G_n$, given by
    \begin{align}\label{eq:Sensing_channel}
    \mathbf G_n =& \sqrt{\frac{\kappa}{1+\kappa}} \sum_{q=1}^{Q} \mathbf G^{\rm los}_{n}(\theta_q^{\rm ta}, \phi_q^{\rm ta}) \nonumber \\ & + \sqrt{\frac{1}{1+\kappa}} \frac{1}{\sqrt{C}} \sum_{c=1}^{C} \mathbf G_{n}^{\rm clu}(\theta^{\rm clu}_c, \phi^{\rm clu}_c),
    \end{align}
    where $\kappa$ is the Rician $K$-factor and $C$ is the number of clutter scatterers. Each clutter-induced NLoS component is modeled as
    \begin{align}\label{eq:clutter_channel}
    \mathbf G_{n}^{\rm clu}(\theta_c, \phi_c) = \alpha^{\rm clu}_c \eta_c \mathbf a^{\sf H}(\theta^{\rm clu}_c, \phi^{\rm clu}_c, \tilde f_n)  \mathbf a(\theta^{\rm clu}_c, \phi^{\rm clu}_c, \tilde f_n),
    \end{align}
    where $\alpha^{\rm clu}_c = \sqrt{\frac{\lambda^2 M^2 \sigma_{\rm clu}}{(4\pi)^3 (l^{\rm clu}_c)^4}}$ with complex small-scale fading $\eta_c \sim \mathcal{CN}(0,1)$ and clutter RCS $\sigma_{\rm clu}$ following the Swerling I model \cite{YOLO}. Here, $l^{\rm clu}_c$ denotes the BS–clutter distance.
    Similarly, let $\theta_k^{\rm cu}$ and $\phi_k^{\rm cu}$ be the elevation and azimuth angles of the $k$-th CU, respectively. A one-way communication channel between the BS and the $k$-th CU on subcarrier $n$ is given by
    \begin{align}\label{eq:Comm_channel}
        \mathbf{h}_{n} (\theta_k^{\rm cu}, \phi_k^{\rm cu}) = \beta_k e^{-j2\pi l_k^{\rm cu}/\lambda} \mathbf a(\theta_k^{\rm cu}, \phi_k^{\rm cu}, \tilde{f_n}),
    \end{align}
    where $\beta_k = \frac{\sqrt{\lambda^2 M}}{4 \pi l_k^{\rm cu}}$ is a communication channel attenuation gain, and $l_k^{\rm cu} = H / \cos \theta_k^{\rm cu}$ represents the distance to the $k$-th CU. Note that both the sensing and communication channels vary across subcarriers because of the beam-squint effect.

    \subsection{Signal Model}
    % \subsection{Signal and Channel Model}\label{Sec:Signal_channel_model}
    In this work, we establish an ISAC transmission frame consisting of $I+1$ sequential stages, with stage index $i$ ranging from $0$ to $I$.
    %ensuring the systematic coverage of $Q$ targets within the ROI. 
    Each stage consists of $T_i$ OFDM symbols, indexed by $t$.  
    The communication signal for the $k$-th CU on subcarrier $n$ of the $t$-th OFDM symbol in stage $i$ is expressed as 
    \begin{align}\label{eq:Data_symbol_vector}
        \mathbf x_{i,k,n} (t) = \mathbf{w}_{k,n} x_{i,k,n}(t),~\forall t\in\{1,\ldots,T_i\},
    \end{align}
    where $\mathbf{w}_{k,n}$ denotes the communication beamforming vector, and $x_{i,k,n} (t)$ represents the communication symbol with power $p_{i,k,n}^{\rm c}$. Note that for $\mathbf{w}_{k,n}$, we omit the indices $i$ and $t$ because the communication strategy remains unchanged during the entire sensing procedure.
    %\textcolor{blue}{ Note that in our strategy, $\mathbf{w}_{i,k,n}(t)$ does not change during the entire sensing procedure because the communication strategy remains unchanged. Therefore, we can simplify the notation to $\mathbf{w}_{k,n}$.} 
    % \textcolor{red}{which is expressed as $x_{i,k,n} (t) = \sqrt{\xi_{i,k,n} (t)} d_{i,k,n} (t)$.}
    The sensing signal on subcarrier $n$ of the $t$-th OFDM symbol in stage $i$ is expressed as 
    \begin{align}\label{eq:Sensing_symbol_vector}
        \mathbf u_{i,n}(t) = \mathbf b_{i,n} u_{i,n}(t),~\forall t\in\{1,\ldots,T_i\},
    \end{align}
    where $\mathbf{b}_{i,n}$ denotes the sensing beamforming vector, with the index $t$ omitted for the same reason as $\mathbf{w}_{k,n}$, and $u_{i,n}(t)$ represents the sensing symbol with power $p_{i,n}^{\rm s}$.\footnote{This framework allows flexible symbol structures where, for instance, communication symbols follow standard schemes like quadrature amplitude modulation and sensing symbols can be Zadoff-Chu sequences with allocated power $p^{\rm s}_{i,n}$. Note that the orthogonality of OFDM guarantees that the system remains free from interference between different subcarriers.}

    At the $k$-th CU, the received signal on subcarrier $n$ of the $t$-th OFDM symbol in stage $i$ is given by 
    \begin{align}\label{eq:Received_comm_signal}
        y_{i,k,n} (t)\! =  &\underbrace {\mathbf{h}_{n} (\theta_k^{\rm cu}, \phi_k^{\rm cu}) \mathbf x_{i,k,n} (t)}_{\text {Desired signal}} \! +\! \underbrace {\mathbf{h}_{n} (\theta_k^{\rm cu}, \phi_k^{\rm cu}) \sum _{l=1,l \neq k}^{K} \mathbf x_{i,l,n} (t)}_{\text {Multiuser interference}} \nonumber \\ & +  \underbrace {\mathbf{h}_{n} (\theta_k^{\rm cu}, \phi_k^{\rm cu}) \mathbf u_{i,n}(t)}_{\text {Sensing signal}} +\, v_{i,k,n} (t),
    \end{align}
    where $v_{i,k,n} (t)$ is the additive white Gaussian noise (AWGN) modeled as a circularly symmetric zero-mean Gaussian random variable with variance $\sigma^2_{k}$.
    In the above expression, the first term represents the desired communication signal for $k$-th CU, the second term accounts for multiuser interference, and the third term denotes interference from the sensing signal.

    At the BS, the same sensing beamforming vector is used for both transmitting and receiving echo signals from the targets. Under this scenario, the echo signal on subcarrier $n$ of the $t$-th OFDM symbol in stage $i$ is expressed as 
    \begin{align}\label{eq:Received_sensing_signal}
        r_{i,n}(t)  = \mathbf b^{\sf H}_{i,n} \mathbf G_{n} \mathbf u_{i,n}(t) + \mathbf{b}_{i,n}^{\sf H} \mathbf v_{i,n} (t),
    \end{align}
    where  $\mathbf v_{i,n} (t)$ is the AWGN vector with variance $\sigma^2_{\rm BS}$. %are i.i.d. circularly symmetric zero-mean Gaussian random variables with zero mean and
    In the above expression, the contribution from CUs is ignored as their significantly lower RCS allows them to be treated as negligible noise or part of the ambient clutter. This is because a user's RCS is typically smaller than a target's RCS, resulting in a significantly weaker reflected signal magnitude \cite{Interference}.
    %In the above expression, the first term represents the reflected echo signal from all targets to the BS, the second term accounts for interference from the communication signal, where $\mathbf G_{n} (\theta_k, \phi_k)$ denotes the round-trip channel on subcarrier $n$ reflected from the $k$-th CU back to the BS. In general, the RCS of the user is significantly smaller compared to the RCS of the target, resulting in a much weaker signal magnitude relative to the reflected echo from the target. The third term represents the projection of the AWGN onto the receive beamforming vector $\mathbf{b}_{i,n}$. % vector $ \mathbf v_{i,n} (t)$, whose entries are i.i.d. circularly symmetric zero-mean Gaussian random variables with zero mean and variance $\sigma^2_v$, 
    % According to [X], it is asymptotically orthogonal to the steering vector at other angles in massive MIMO systems. 

    \begin{figure}[t]
        \centering %\vspace{-3mm}s
        {\epsfig{file=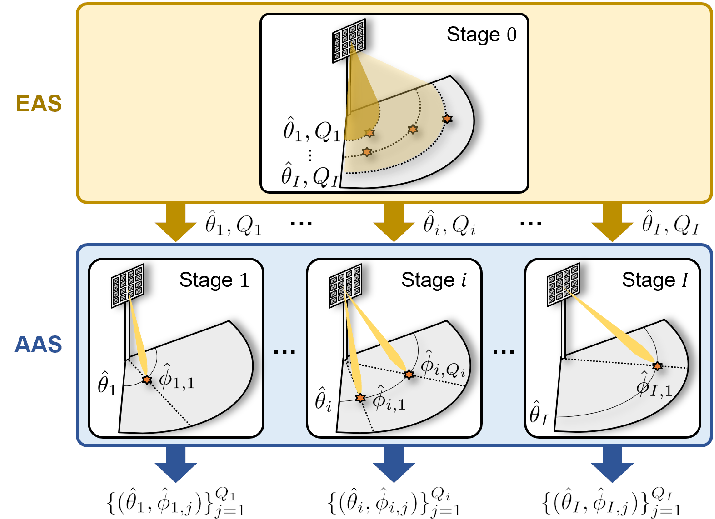,width=7cm}}%\vspace{-3mm}
        \caption{A framework of the proposed hierarchical sensing framework with two processes: Elevation angle sensing (EAS) and azimuth angle sensing (AAS).}\label{fig:sensing_framework}  %\vspace{-3mm}
    \end{figure}
    %An illustration of an ISAC system where the BS equipped with UPA  communicates with $K$ single-antenna CUs and senses $Q$ passive targets.
    \section{Beamforming Design for the \\  Proposed Hierarchical Sensing Framework}\label{Sec:Hierarchical} 
    In this work, we propose a hierarchical sensing framework that enables efficient estimation of the elevation and azimuth angles of the targets by exploiting the beam-squint effect through a structured, multi-stage sensing process.
    %to efficiently cover the ROI with only a few beam scans.
    % Specifically, the proposed framework comprises two processes.  The first process, corresponding to stage $0$, is dedicated to estimating the elevation angles of the targets, while the second process, spanning stages $1$ through $I$, focuses on estimating the azimuth angles of the targets. At each stage, the beam-squint effect is exploited to cover $N$ elevation or azimuth grid points by controlling both TTD lines and PSs of the UPA. 
    %Based on this idea, in what follows, we present the sensing beamforming vector design for both processes in the proposed framework. We then describe the design of the communication beamforming vector within the proposed sensing framework.
    
    \subsection{Overview of the Proposed Framework}
    The proposed sensing framework consists of two sequential processes: (i) elevation angle sensing (EAS) and (ii) azimuth angle sensing (AAS), as illustrated in Fig.~\ref{fig:sensing_framework}. In both processes, the beam-squint effect is exploited to cover $N$ elevation or azimuth grid points by jointly controlling the TTD lines and PSs of the UPA, where $N$ corresponds to the previously defined number of subcarriers. The high-level procedures of each process are summarized as follows:
    \begin{itemize}
        \item {\bf EAS Process:} This is a single-stage process (referred to as stage $0$) dedicated to estimating the elevation angles of $Q$ targets. As a result, $I \leq Q$ elevation angles in the ROI are estimated, denoted by $\hat{\theta_1},\ldots,\hat{\theta_I}$. The value $I$ can be smaller than $Q$ when several targets are located near the same elevation angle grid. The EAS process also determines the number of targets, denoted by $Q_i$ (with $Q_i\leq Q$), to be estimated at each stage of the subsequent AAS process for each elevation angle $\hat{\theta}_i$. For notational consistency, we define $Q_0 = Q$. 

        \item {\bf AAS Process:} This is a multi-stage process, with each stage $i\in\{1,\ldots,I\}$ corresponding to one of the estimated elevation angles from the EAS process. At each stage $i$, the azimuth angles of the $Q_i$ targets are estimated for the fixed elevation angle $\hat{\theta}_i$. The estimated azimuth angles in stage $i$ are denoted by $\hat{\phi}_{i,1},\ldots,\hat{\phi}_{i,Q_i}$. Consequently, the positions of the $Q_i$ targets are determined as $(\hat{\theta}_{i},\hat{\phi}_{i,1}),\ldots,(\hat{\theta}_{i},\hat{\phi}_{i,Q_i})$.
    \end{itemize}

    A key enabler of this framework is the sensing beamforming design, which enables a single beam scan to simultaneously cover $N$ elevation or azimuth grid points by utilizing the beam-squint effect. The sensing beamforming vector designs for the EAS and AAS processes are presented in Sec.~\ref{Sec:Sensing_BF_EAS} and Sec.~\ref{Sec:Sensing_BF_AAS}, respectively. The design of the communication beamforming vector within the proposed framework is described in Sec.~\ref{Sec:Comm_BF}.

    \subsection{Sensing Beamforming for EAS Process}\label{Sec:Sensing_BF_EAS}
    The goal of the EAS process, corresponding to stage $0$, is to estimate the elevation angles of targets within the ROI.
    To achieve this goal, each subcarrier beamforming direction in stage $0$ is designed to sweep from the minimum elevation angle $\theta_{\rm min}$ to the maximum elevation angle $\theta_{\rm max}$ as the frequency $\tilde{f}_n$ increases from $0$ to $F$. Accordingly, the first subcarrier is assigned to $\theta_{\rm min}$ and the last subcarrier to $\theta_{\rm max}$. Leveraging the frequency-dependent deviations induced by the beam-squint effect, this approach enables the simultaneous sensing of multiple elevation angles between $\theta_{\rm min}$ and $\theta_{\rm max}$ across subcarriers.
    By optimizing the beamforming parameters to maximize the array gain at these two points, the intermediate grid points for the remaining $N-2$ subcarriers are inherently determined by the beam-squint effect \cite{beam_squint_split}.

    To determine the sensing beamforming vector for stage $0$, we first express a sensing beamforming vector on subcarrier $n$ as a function of the PSs connected to the sensing RF chain and TTD parameters.
    Let $(\theta_{0}^{\rm b}, \phi_{0}^{\rm b})$ be the elevation and azimuth angles for the PSs in stage $0$. 
    The time-domain response of the TTD line at the UPA antenna element position $(m_{\rm h}, m_{\rm v})$ is expressed as $\delta(t - t^{\rm s}_{0,(m_{\rm h},m_{\rm v})})$, where $t^{\rm s}_{0,(m_{\rm h},m_{\rm v})}$ represents the time delay for sensing induced by the TTD unit in stage $0$. The frequency-domain response of the sensing analog beamforming vector on subcarrier $n$, aimed at a specific beamforming direction characterized by the angular parameters $(\theta_{0}^{\rm b}, \phi_{0}^{\rm b})$, is given by
    \begin{align}\label{eq:Sensing_BF_stage0}
        \mathbf b_{0,n} = \underbrace { {\sf diag}({\bf t}^{\rm s}_{0,n}) }_{\text {TTD response}} \underbrace { \mathbf a^{\sf H}(\theta_0^{\rm b}, \phi_0^{\rm b},0) }_{\text {PS response}},
    \end{align}
    where ${\bf t}^{\rm s}_{0,n}=\big [e^{-j2\pi\tilde{f}_n t^{\rm s}_{0,(1,1)} }, \cdots, e^{-j2\pi\tilde{f}_n t^{\rm s}_{0,(M_{\rm h},M_{\rm v})} } \big]$ and $\mathbf a(\theta_0^{\rm b}, \phi_0^{\rm b},0)$ is generally the steering vector for the desired direction across all subcarriers under the absence of beam‑squint, but now serves as the steering vector pointing toward the starting point of the beam‑squint effect. Note that $\mathbf b_{0,n}$ captures the combined effects of PS and TTD on the frequency-dependent beamforming response.

    Let $(\vartheta_n, \varphi_n)$ be the grid point for the elevation and azimuth angle steered by subcarrier $n$. 
    The array gain of the sensing beamforming vector $\mathbf b_{0,n}$ in \eqref{eq:Sensing_BF_stage0} and the steering vector $\mathbf a(\vartheta_n, \varphi_n, \tilde{f}_n)$  is given as follows \cite{beam_squint_split}:
    % \begin{align}\label{eq:array_gain_stage0}
    %     &|\mathbf b^{\sf H}_{0,n} \mathbf G_{n}(\vartheta_n, \varphi_n) \mathbf b_{0,n} |\nonumber \\
    %     &= \alpha_n \big|\big ( \mathbf a_{M_{\rm h}}(\vartheta_n, \varphi_n,\tilde{f}_n) \otimes \mathbf a_{M_{\rm v}}(\vartheta_n, \tilde{f}_n)\big ) \times {\sf diag}({\bf t}_0) \nonumber \\ 
    %     &~~~\times \big ( \mathbf a_{M_{\rm h}}(\theta_{0}^{\rm b}, \phi_{0}^{\rm b}, 0) \otimes \mathbf a_{M_{\rm v}}(\theta_{0}^{\rm b}, 0)\big )^{\sf H} \big|^2 \nonumber \\
    %     &=\alpha_n \big| \mathbf a_{M_{\rm h}}(\vartheta_n, \varphi_n,\tilde{f}_n)   {\sf diag}({\bf t}_{0,h}) \mathbf a^{\sf H}_{M_{\rm h}}(\theta_0^{\rm b}, \phi_0^{\rm b}, 0)  \nonumber\\ 
    %     &~~~\otimes \mathbf a_{M_{\rm v}}(\vartheta_n, \tilde{f}_n)   {\sf diag}({\bf t}_{0,v}) \mathbf a^{\sf H}_{M_{\rm v}}(\theta_0^{\rm b}, 0)\big|^2.
    % \end{align}
    \begin{align}\label{eq:array_gain_stage0}
        & \big |\mathbf a(\vartheta_n, \varphi_n, \tilde{f}_n) \mathbf b_{0,n} \big |\nonumber \\
        &= \big| \mathbf a(\vartheta_n, \varphi_n, \tilde{f}_n) {\sf diag}({\bf t}^{\rm s}_{0,n}) \mathbf a^{\sf H}(\theta_0^{\rm b}, \phi_0^{\rm b},0) \big|  \nonumber \\ 
        &= \big| \mathbf a_{\rm h}(\vartheta_n, \varphi_n,\tilde{f}_n) \otimes \mathbf a_{\rm v}(\vartheta_n, \tilde{f}_n) \times {\sf diag}({\bf t}^{\rm s}_{0,n}) \nonumber\\ 
        &~~~\times \mathbf a^{\sf H}_{\rm h}(\theta_0^{\rm b}, \phi_0^{\rm b}, 0) \otimes \mathbf a^{\sf H}_{\rm v}(\theta_0^{\rm b}, 0) \big|.
    \end{align}
    By exploiting the mixed-product property of the Kronecker product with $t^{\rm s}_{0,(m_{\rm h},m_{\rm v})}=t^{\rm s,h}_{0,(m_{\rm h})}+t^{\rm s,v}_{0,(m_{\rm v})}$, the array gain is rewritten as
    \begin{align}\label{eq:revised_array_gain_stage0}
        &\big |\mathbf a(\vartheta_n, \varphi_n, \tilde{f}_n) \mathbf b_{0,n} \big |\nonumber \\
        &=\big| \big ( \mathbf a_{\rm h}(\vartheta_n, \varphi_n,\tilde{f}_n) {\sf diag}({\bf t}^{\rm s,h}_{0,n}) \mathbf a^{\sf H}_{\rm h}(\theta_0^{\rm b}, \phi_0^{\rm b} ,0) \big ) \nonumber\\ 
        &~~~\times \big (\mathbf a_{\rm v}(\vartheta_n, \tilde{f}_n) {\sf diag}({\bf t}^{\rm s,v}_{0,n}) \mathbf a^{\sf H}_{\rm v}(\theta_0^{\rm b}, 0)\big ) \big|,
    \end{align}
    where ${\bf t}^{\rm s,h}_{0,n} = \big [e^{-j2\pi\tilde{f}_n t^{\rm s,h}_{0,(1)} }, \cdots, e^{-j2\pi\tilde{f}_n t^{\rm s,h}_{0,(M_{\rm h})} } \big]$ and ${\bf t}^{\rm s,v}_{0,n} = \big [e^{-j2\pi\tilde{f}_n t^{\rm s,v}_{0,(1)} }, \cdots, e^{-j2\pi\tilde{f}_n t^{\rm s,v}_{0,(M_{\rm v})} } \big]$.

    As mentioned earlier, our goal is to determine $\theta_0^{\rm b}$, $\phi_0^{\rm b}$, $\{t^{\rm s,h}_{0,(m_{\rm h})}\}_{m_{\rm h}}$, and $\{t^{\rm s,v}_{0,(m_{\rm v})}\}_{m_{\rm v}}$ to maximize the array gain in \eqref{eq:revised_array_gain_stage0} at the first and last subcarriers. Note that both the horizontal and vertical steering vectors inherently incorporate the elevation angle $\vartheta_n$, as can be seen in \eqref{eq:vectorized_steering_vector}.
    %both the horizontal and vertical components of the array gain
    Consequently, when considering the beam-squint effect, the maximization value of the array gain with respect to $\vartheta_n$ becomes non-unique, leading to an inherent ambiguity in the optimization process. To address this issue, we assume that the horizontal beam pattern across the predefined azimuth range is approximately a constant, which can be approached by generating a nearly flat beam pattern by adjusting $\phi_0^{\rm b}$ and $\{t^{\rm s,h}_{0,(m_{\rm h})}\}_{m_{\rm h}}$ based on the existing algorithms in \cite{beam_broad1,beam_broad2}.
    % \textcolor{blue}{To address this issue, a wide-beam technique enables the horizontal component in \eqref{eq:revised_array_gain_stage0}, involving $\phi_0^{\rm b}$ and $\{t_{0,h}^{(m_{\rm h})}\}_{m_{\rm h}}$, to be approximated as a constant. Specifically, this assumption is useful because the beam pattern across the predefined azimuth range approaches a flat beam pattern by employing algorithms proposed in \cite{beam_broad1,beam_broad2}.}
    In \eqref{eq:M_h_steering_vector}, the term $\sin\theta \cos\phi$ characterizes the composite angle of departure (AoD), denoted by $\psi$, which is a function of the elevation and azimuth angles and is defined as $\psi = \arccos(\sin\theta \cos\phi)$. 
    Then, the minimum and maximum composite AoDs needed to cover the entire ROI, accounting for the beam-squint effect, are determined by $\psi_{\text{min}} = \arccos(\sin \theta_{\text{max}} \cos \phi_{\text{min}})$ and $\psi_{\text{max}} = \arccos \big(\sin \theta_{\text{max}} \cos \phi_{\text{max}} \big(1+\frac{F}{f_c}\big) \big)$, respectively. Following \cite{ideal_beam}, the integral of the square of the horizontal array gain over the interval [$\psi_{\text{min}}$, $\psi_{\text{max}}$) can be assumed to be $\frac {2 \pi}{M_{\rm h}}$.
    % assumed to satisfy
    % \begin{align}\label{eq:int_ideal_beam_pattern}
    %     \int _{\psi_{\text{min}}}^{\psi_{\text{max}}} \! \underbrace {\Big | \big ( \mathbf a_{M_{\rm h}}(\vartheta_n, \varphi_n,\tilde{f}_n)   {\sf diag}({\bf t}_{0,n,h})  \mathbf a^{\sf H}_{M_{\rm h}}(\theta_0^{\rm b}, \phi_0^{\rm b},0) \big ) \Big |^2 }_{\text {Constant}}\! d\psi \!=\!\frac {2 \pi}{M_{\rm h}}.
    % \end{align}
    % Using \eqref{eq:int_ideal_beam_pattern}, 
    %This implies that, we can derive the constant value of the ideal beam pattern, which can then be utilized to compute the array gain in  \eqref{eq:revised_array_gain_stage0}.
    Consequently, the horizontal array gain in \eqref{eq:revised_array_gain_stage0} can be approximated as
    \begin{align}\label{eq:ideal_beam_pattern}
             &\Big | \mathbf a_{\rm h}(\vartheta_n, \varphi_n,\tilde{f}_n) {\sf diag}({\bf t}^{\rm s,h}_{0,n})  \mathbf a^{\sf H}_{\rm h}(\theta_0^{\rm b}, \phi_0^{\rm b},0) \Big | \nonumber \\ &= \sqrt{\frac {2 \pi}{M_{\rm h}\big (\psi_{\text{max}} - \psi_{\text{min}} \big )} },~\forall \psi \in [\psi_{\text{min}},\psi_{\text{max}}).
    \end{align}
    Now, the remaining term in \eqref{eq:revised_array_gain_stage0} is the vertical array term that depends on $\theta_0^{\rm b}$ and  $\{t^{\rm s,v}_{0,(m_{\rm v})}\}_{m_{\rm v}}$. To determine these values, we express the vertical array term in summation form, following the approach in \cite{beam_squint_split}, given by
    % \begin{align}\label{eq:stage1_vertical_array_gain}
    %     &\mathbf a_{M_{\rm v}}(\vartheta_n, \tilde{f}_n)   {\sf diag}({\bf t}_{0,v}) \mathbf a^{\sf H}_{M_{\rm v}}(\theta_0^{\rm b}, 0) \nonumber \\
    %     &=\frac{1}{M_{\rm v}}\sum^{M_{\rm v}}_{m_{\rm v} = 1} e^{-j\pi (m_{\rm v} - 1) \cos \vartheta_n \big ( 1+\frac{\tilde{f}_n}{f_c} \big )}  \nonumber \\ 
    %     &~~~\times e^{j\pi (m_{\rm v} - 1)\cos \theta_0^{\rm b}} e^{-j2\pi \tilde{f}_n t_{0,m_{\rm v}}}. \\
    %     &= \frac{1}{M_{\rm v}}\sum^{M_{\rm v}}_{m_{\rm v} = 1} e^{-j\pi \big ( (m_{\rm v}-1) (\cos \vartheta_n \big ( 1+\frac{\tilde{f}_n}{f_c} \big ) -\cos  \theta_0^{\rm b} ) + 2 \tilde{f}_n t_{0,m_{\rm v}} \big )}
    % \end{align}
    \begin{align}\label{eq:stage0_vertical_array_gain}
        \mathbf a_{\rm v}(\vartheta_n, \tilde{f}_n)   {\sf diag}({\bf t}^{\rm s,v}_{0,n}) \mathbf a^{\sf H}_{\rm v}(\theta_0^{\rm b}, 0) = \frac{1}{M_{\rm v}}\sum^{M_{\rm v}}_{m_{\rm v} = 1} e^{-j\pi g_{0,n}(m_{\rm v}) },
        %e^{-j\pi \big ( (m_{\rm v}-1) \big (\cos \vartheta_n \big ( 1+\frac{\tilde{f}_n}{f_c} \big ) -\cos  \theta_0^{\rm b} \big ) + 2 \tilde{f}_n t^{(m_{\rm v})}_{0} \big )},
    \end{align}
    where
    \begin{align}\label{eq:exponent_term_stage0}
         g_{0,n}(m_{\rm v}) \!=\! \big(m_{\rm v} \! - \! 1 \big) \! \bigg(\! \cos \vartheta_n \Big ( 1 \!+ \! \frac{\tilde{f}_n}{f_c} \Big ) \!-\! \cos  \theta_0^{\rm b}  \bigg) \!+\! 2 \tilde{f}_n t^{\rm s,v}_{0,(m_{\rm v})}.
    \end{align}    
    % where a key part of the exponent term, denoted as $g_{0,v}(n)$, is derived to be zero in order to maximize \eqref{eq:stage0_vertical_array_gain}. $g_{0,v}(n)$ is obtained as
    % \begin{align}\label{eq:exponent_term_stage0}
    %      g_{0,v}(n) = \big(m_{\rm v}-1 \big) \bigg (\cos \vartheta_n \Big ( 1+\frac{\tilde{f}_n}{f_c} \Big ) -\cos  \theta_0^{\rm b}  \bigg ) + 2 \tilde{f}_n t^{(m_{\rm v})}_{0} .
    % \end{align}
    Since the frequency deviation of the first subcarrier is given by $\tilde{f}_1=0$, the exponent term $g_{0,1}(m_{\rm v})$ for the first subcarrier is expressed as
    \begin{align}\label{eq:exponent_term_stage0_first}
         g_{0,1}(m_{\rm v}) = \big(m_{\rm v}-1 \big) \big (\cos \vartheta_1 -\cos  \theta_0^{\rm b}  \big ).
    \end{align}
    To ensure that the sensing beamforming vector of the first subcarrier has the maximum gain at the minimum elevation angle $\theta_{\text{min}}$, we set $\vartheta_1 = \theta_{\text{min}}$ and determine $\theta_{0}^b = \theta_{\text{min}}$. %This choice ensures that $g_{0,1}(m_{\rm v}) = 0$ for all $m_{\rm v} \in\{1,\ldots,M_{\rm v}\}$. 
    % As previously mentioned, the first subcarrier is aligned with $\theta_{\text{min}}$ and the last subcarrier with $\theta_{\text{max}}$ to account for the beam squint effect, thereby fixing $\vartheta_1 = \theta_{\text{min}}$ and $\vartheta_N = \theta_{\text{max}}$. Since $\tilde{f}_1=0$ represents the frequency deviation of the first subcarrier, the exponent term $g_{0,1}(m_{\rm v})$ for the first subcarrier is given by 
    % \begin{align}\label{eq:exponent_term_stage0_first}
    %      g_{0,1}(m_{\rm v}) = \big(m_{\rm v}-1 \big) \big (\cos \vartheta_1 -\cos  \theta_0^{\rm b}  \big ).
    % \end{align}
    % Hence, the optimal value of $\theta_{0}^b$ that ensures $g_{0,v}(1) = 0$ is determined as $\theta_{0}^b = \vartheta_1$. 
    Similarly, from $\theta_{0}^b = \theta_{\text{min}}$ and $\tilde{f}_N=F$, the exponent term $g_{0,N}(m_{\rm v})$ for the last subcarrier (i.e., the $N$-th subcarrier) is expressed as 
    \begin{align}\label{eq:exponent_term_stage0_last}
         g_{0,N}(m_{\rm v}) \! = \! \big(m_{\rm v} \! - \! 1 \big) \! \bigg (\! \cos \vartheta_N \Big ( 1 \! + \! \frac{F}{f_c} \Big ) \! - \! \cos \theta_{\text{min}} \! \bigg ) \!\! + \! 2 F t^{\rm s,v}_{0,(m_{\rm v})}.
    \end{align}
    To ensure that the sensing beamforming vector of the last subcarrier has the maximum gain at the maximum elevation angle $\theta_{\text{max}}$, we set $\vartheta_N = \theta_{\text{max}}$. 
    Then, the optimal value\footnote{While the TTD-selection rule is applicable even without beam squint by maximizing gain at boundary angles like $\theta_{\rm min}$ and $\theta_{\rm max}$, a non-negligible squint is practically preferable for both EAS and AAS as it stabilizes the TTD range and improves hardware robustness.} of $t^{\rm s,v}_{0,(m_{\rm v})}$ that ensures $g_{0,N}(m_{\rm v}) = 0$ is  determined as
   \begin{align}\label{eq:TTD_mv_stage0}
         t^{\rm s,v}_{0,(m_{\rm v})} = \frac{(m_{\rm v} - 1)(\cos \theta_{\text{min}} - \cos\theta_{\text{max}} (1+ \frac{F}{f_c}))}{2F},
    \end{align}
    for all $m_{\rm v} \in\{1,\ldots,M_{\rm v}\}$.
    %  remark
    % \textcolor{blue}{{\bf Remark 1 (Applicability when beam squint is negligible):} The TTD-selection rule remains applicable even when the beam-squint effect is negligible. The TTD values can still be determined by maximizing the array gain at the start angle $\theta_{\rm min}$ and the end angle $\theta_{\rm max}$, which implicitly fixes the intermediate subcarrier-angle mapping. Nevertheless, in practice, a non-negligible beam squint is preferable because it stabilizes the admissible TTD range across subcarriers and improves robustness to hardware quantization and calibration errors. This principle similarly applies to the AAS process, where the TTDs are determined based on the angles $\phi_{\rm min}$ and $\phi_{\rm max}$.}
    Finally, by substituting the determined values of $\theta_{0}^b$ and $t^{\rm s,v}_{0,(m_{\rm v})}$ back into \eqref{eq:exponent_term_stage0} and solving for $\vartheta_n$ that satisfies $g_{0,n}(m_{\rm v}) = 0$, the elevation angle of subcarrier $n$ due to the beam-squint effect is obtained as
    \begin{align}\label{eq:vartheta_n}
        \vartheta_{n} {=} \arccos \!\left (\! \frac { \cos \theta_{\text{min}}- \frac{\tilde{f}_n}{F} \Big ( \cos \theta_{\text{min}}  - \cos \theta_{\text{max}} \big ({1 + \frac {F}{f_{c}}}\big ) \Big ) }{1+\frac {\tilde{f}_n}{f_{c}} } \right).
    \end{align} 
    A prominent feature of our sensing beamforming vector ${\bf b}_{0,n}$ in stage $0$ is that it concurrently covers $N$ different elevation angles in the ROI, given by $\vartheta_{1},\ldots, \vartheta_{N}$, treated as grids, using $N$ subcarriers. This differs from a conventional sensing approach, which requires transmitting $N$ different beams to cover $N$ grid points. Using the beam-squint effect, our beamforming vector significantly reduces the sensing time required to scan and determine the elevation angles of the targets. Details of the target detection strategy for stage $0$ will be described in Sec.~\ref{Sec:Detection}.

    \subsection{Sensing Beamforming for AAS Process}\label{Sec:Sensing_BF_AAS}
    The goal of stage $i$ in the AAS process is to estimate the azimuth angles of the $Q_i$ targets for a given elevation angle $\hat{\theta}_i$. To accomplish this goal,  the sensing beamforming direction in stage $i$ is designed to scan from the minimum azimuth angle $\phi_{\rm min}$ to the maximum azimuth angle $\phi_{\rm max}$ in the ROI as the frequency deviation $\tilde{f}_n$ increases from $0$ to $F$. Then the beam-squint effect inherently determines the $N-2$ intermediate grid points, given that the first and last subcarriers are allocated to $\phi_{\rm min}$ and $\phi_{\rm max}$, respectively, with beamforming parameters optimized to maximize the array gain at these two points.
    
    % At stage $0$, the detection strategy described in Sec.~\ref{Sec:Target_detection} identifies $I$ potential elevation angles $\theta_i,~i=1,2,\ldots,I$,  which are elements of the set  $\{ \vartheta_{1}, \vartheta_{2}, \ldots, \vartheta_{N} \}$  covering $Q$ targets. We define $Q_0$ as the number of targets to be identified at stage $0$, and since all elevation angles are estimated at a single stage, $Q_0$ can be considered as $Q$. In the subsequent stage $i$, $Q_i$ azimuth angles are estimated where targets are anticipated at the given elevation angle $\theta_i$. The total number of estimated targets, obtained through all stages $i,~i=1,2,\ldots, I$, is given by $Q_1 + Q_2 + \cdots + Q_I = Q$. 

    %Consisting of $I$ stages, the beam squint effect is used to estimate the azimuth angle, where targets are expected to be located at each stage. Similar to stage $0$, the subcarrier beamforming direction is intended to transition from the initial angle $\phi_{\rm min}$ to the final angle $\phi_{\rm max}$ as the frequency deviation $\tilde{f}_n$ increases from $0$ to $F$. The beam squint effect inherently determines the \(N-2\) intermediate grid points, given that the first and last subcarriers are allocated to \(\phi_{\rm min}\) and \(\phi_{\rm max}\), respectively, with beamforming parameters optimized to maximize the array gain at these two points.

    To design the sensing beamforming vector in stage $i$, we express a sensing beamforming vector on subcarrier $n$ as a function of the PSs connected to the sensing RF chain and TTD line parameters. % $\mathbf{b}_{i,n}$ is also expressed as a function of the PS and TTD parameters. 
    Unlike in stage $0$, the elevation angle of the PS is fixed as $\hat{\theta}_i$, while the azimuth angle is set to a programmable parameter $\phi^b_i$. That is, in stage $i$, the parameter of the PS is determined as $(\hat{\theta}_{i}, \phi_{i}^{\rm b})$, implying that we only need to determine the azimuth angle $ \phi_{i}^{\rm b}$ for each stage. Meanwhile, the beam-squint effect is controlled using TTD lines, where the TTD unit at the antenna element position $(m_{\rm h}, m_{\rm v})$ induces a time delay $t^{\rm s}_{i,(m_{\rm h},m_{\rm v})}$. The frequency-domain response of the sensing beamforming vector for subcarrier $n$ in stage $i$ is expressed as
    \begin{align}\label{eq:Sensing_BF_stage_i}
        \mathbf b_{i,n} =  {\sf diag}({\bf t}^{\rm s}_{i,n}) \mathbf a^{\sf H}(\hat{\theta}_{i}, \phi_i^{\rm b},0),
    \end{align}
    where ${\bf t}^{\rm s}_{i,n} = \big [e^{-j2\pi\tilde{f}_n t^{\rm s}_{i,(1,1)} },  \cdots, e^{-j2\pi\tilde{f}_n t^{\rm s}_{i,(M_{\rm h},M_{\rm v})} } \big]$. %and this formulation closely resembles \eqref{eq:Sensing_BF_stage0} with the specified elevation angle. 
    Let $(\hat{\theta}_{i}, \varphi_n)$ be the grid point for the azimuth angle steered by subcarrier $n$ in stage $i$. The array gain of the sensing beamforming vector $\mathbf b_{i,n}$ in \eqref{eq:Sensing_BF_stage_i} and the steering vector $\mathbf a(\hat{\theta}_{i}, \varphi_n, \tilde{f}_n)$ is 
    \begin{align}\label{eq:revised_array_gain_stage_i}
        & \big |\mathbf a(\hat{\theta}_{i}, \varphi_n, \tilde{f}_n) \mathbf b_{i,n} \big |\nonumber \\
        &= \big| \mathbf a_{\rm h}(\hat{\theta}_{i}, \varphi_n,\tilde{f}_n) \otimes \mathbf a_{\rm v}(\hat{\theta}_{i}, \tilde{f}_n) \times {\sf diag}({\bf t}^{\rm s}_{i,n}) \nonumber\\ 
        &~~~\times \mathbf a^{\sf H}_{\rm h}(\hat{\theta}_{i}, \phi_i^{\rm b}, 0) \otimes \mathbf a^{\sf H}_{\rm v}(\hat{\theta}_{i}, 0) \big| %\label{eq:array_gain_stage_i} \\  
        \nonumber \\
        &\overset{(a)}{=}\big| \big ( \mathbf a_{\rm h}(\hat{\theta}_{i}, \varphi_n,\tilde{f}_n) {\sf diag}({\bf t}^{\rm s,h}_{i,n}) \mathbf a^{\sf H}_{\rm h}(\hat{\theta}_{i}, \phi_i^{\rm b} ,0) \big ) \nonumber\\ 
        &~~~\times \big (\mathbf a_{\rm v}(\hat{\theta}_{i}, \tilde{f}_n) {\sf diag}({\bf t}^{\rm s,v}_{i,n}) \mathbf a^{\sf H}_{\rm v}(\hat{\theta}_{i}, 0)\big ) \big|, 
    \end{align}
    where $(a)$ holds due to the property of the Kronecker product, specifically when $t^{\rm s}_{i,(m_{\rm h},m_{\rm v})}=t^{\rm s,h}_{i,(m_{\rm h})}+t^{\rm s,v}_{i,(m_{\rm v})}$ with ${\bf t}^{\rm s,h}_{i,n} = \big [e^{-j2\pi\tilde{f}_n t^{\rm s,h}_{i,(1)} }, \cdots, e^{-j2\pi\tilde{f}_n t^{\rm s,h}_{i,(M_{\rm h})} } \big]$ and ${\bf t}^{\rm s,v}_{i,n} = \big [e^{-j2\pi\tilde{f}_n t^{\rm s,v}_{i,(1)} }, \cdots, e^{-j2\pi\tilde{f}_n t^{\rm s,v}_{i,(M_{\rm v})} } \big]$. %similar to the decomposition in \eqref{eq:revised_array_gain_stage0}.
    %where, as seen in stage $0$, by decomposing \eqref{eq:array_gain_stage_i} into its horizontal and vertical components, similar to the decomposition in \eqref{eq:revised_array_gain_stage0}, it can be expressed as \eqref{eq:revised_array_gain_stage_i}.

    % Unlike stage $0$, where the elevation angle was not uniquely determined due to the beam squint effect, the elevation angle $\theta$ in this case is fixed at $\theta_i$. Since only the horizontal component steering vector is associated with $\varphi_n$, as shown in \eqref{eq:revised_array_gain_stage_i}, the azimuth angle can therefore be uniquely determined by leveraging the beam squint effect. Consequently, $\phi_i^{\rm b}$, ${\bf t}_{i,h}$, and ${\bf t}_{i,v}$ are determined to maximize the horizontal and vertical array gain terms, respectively.
    Our goal is to determine $\phi_i^{\rm b}$, $\{t^{\rm s,h}_{i,(m_{\rm h})}\}_{m_{\rm h}}$, and $\{t^{\rm s,v}_{i,(m_{\rm v})}\}_{m_{\rm v}}$ to maximize the array gain in \eqref{eq:revised_array_gain_stage_i} at the first and last subcarriers. To determine $\{t^{\rm s,v}_{i,(m_{\rm v})}\}_{m_{\rm v}}$, we express the vertical array gain in \eqref{eq:revised_array_gain_stage_i} in summation form as
    \begin{align}\label{eq:stage_i_vertical_array_gain}
        &\mathbf a_{\rm v}(\hat{\theta}_i, \tilde{f}_n)   {\sf diag}({\bf t}^{\rm s,v}_{i,n}) \mathbf a^{\sf H}_{\rm v}(\hat{\theta}_i, 0) = \frac{1}{M_{\rm v}}\sum^{M_{\rm v}}_{m_{\rm v} = 1} e^{-j\pi g_{i,n}(m_{\rm v}) },
        %\nonumber \\ &= \frac{1}{M_{\rm v}}\sum^{M_{\rm v}}_{m_{\rm v} = 1} e^{-j\pi \tilde{f}_n \big ( (m_{\rm v}-1) \frac{\cos \theta_i}{f_c} + 2 t^{(m_{\rm v})}_{i} \big )},
    \end{align}
    where %a significant part of the exponent term, denoted as $g_{i,n}(m_{\rm v})$, is determined to be zero to maximize \eqref{eq:stage_i_vertical_array_gain}. The expression for $g_{i,v}(n)$ is given as
    \begin{align}\label{eq:exponent_term_mv_stage_i}
         g_{i,n}(m_{\rm v}) = \tilde{f}_n \left ( \big(m_{\rm v}-1 \big) \frac{\cos \hat{\theta}_i}{f_c} + 2 t^{\rm s,v}_{i,(m_{\rm v})} \right ).
    \end{align}
    The expression in \eqref{eq:exponent_term_mv_stage_i} shows that regardless of $\tilde{f}_n$, the exponent term $g_{i,n}(m_{\rm v})$ can be set to zero by choosing $t^{\rm s,v}_{i,(m_{\rm v})}$ as 
    \begin{align}\label{eq:TTD_mv_stage_i}
         t^{\rm s,v}_{i,(m_{\rm v})} = \frac{(1 - m_{\rm v})\cos\hat{\theta}_i}{2f_c},~\forall m_{\rm v} \in\{1,\ldots,M_{\rm v}\}.
    \end{align}
    Similarly, to determine $\phi_i^{\rm b}$ and $\{t^{\rm s,h}_{i,(m_{\rm h})}\}_{m_{\rm h}}$, we express the horizontal array gain in \eqref{eq:revised_array_gain_stage_i} in summation form:
    \begin{align}\label{eq:stage_i_hirizontal_array_gain}
        & \mathbf a_{\rm h}(\hat{\theta}_{i}, \varphi_n,\tilde{f}_n) {\sf diag}({\bf t}^{\rm s,h}_{i,n}) \mathbf a^{\sf H}_{\rm h}(\hat{\theta}_{i}, \phi_i^{\rm b} ,0) \!=\!  \frac{1}{M_{\rm h}} \! \sum^{M_{\rm h}}_{m_{\rm h} = 1}\!\! e^{-j\pi g_{i,n}(m_{\rm h})},
        %\nonumber \\ &=\! \frac{1}{M_{\rm h}} \!\!\sum^{M_{\rm h}}_{m_{\rm h} = 1}\!\! e^{-j\pi \big ( (m_{\rm h}-1)\sin \theta_i  \big (\cos \varphi_n \big ( 1+\frac{\tilde{f}_n}{f_c} \big ) - \cos \phi^{\rm b}_i \big ) + 2 \tilde{f}_n t^{(m_{\rm h})}_{i} \big )},
    \end{align}
    where %a primary part of the exponent term, represented as $g_{i,h}(n)$, is determined to be zero to achieve the maximum value of \eqref{eq:stage_i_hirizontal_array_gain}. The expression for $g_{i,h}(n)$ is derived as 
    \begin{align}\label{eq:exponent_term_mh_stage_i}
         g_{i,n}(m_{\rm h}) =& \big(m_{\rm h}-1 \big) \sin \hat{\theta}_i \bigg ( \cos \varphi_n \Big ( 1+\frac{\tilde{f}_n}{f_c} \Big ) -\cos  \phi_i^{\rm b} \bigg ) \nonumber \\ &+ 2 \tilde{f}_n t^{\rm s,h}_{i,(m_{\rm h})}.
    \end{align}
    %As discussed earlier, to account for the beam squint effect, the first subcarrier is mapped to $\phi_{\rm min}$ and the last subcarrier to $\phi_{\rm max}$, thereby fixing $\varphi_1 = \phi_{\rm min}$ and $\varphi_N = \phi_{\rm max}$. Based on this, by substituting $\tilde{f}_1=0$ into $\tilde{f}_n$ in \eqref{eq:exponent_term_mh_stage_i}, the value of $\phi^{\rm b}_i$ that maximizes the horizontal array gain for the first subcarrier can be derived as
    From $\tilde{f}_1=0$, the exponent term $g_{i,1}(m_{\rm h})$ for the first subcarrier is given by 
    \begin{align}\label{eq:exponent_term_stage_i_first}
         g_{i,1}(m_{\rm h}) \!=\! \big(m_{\rm h}-1 \big) \sin \hat{\theta}_i \big (\cos \varphi_1 -\cos  \phi^{\rm b}_i  \big ).
    \end{align}
    To ensure that the horizontal array gain of the first subcarrier is maximized at the minimum azimuth angle $\phi_{\text{min}}$, we set $\varphi_1 = \phi_{\text{min}}$ and determine $\phi_{i}^b = \phi_{\text{min}}$.
    Similarly, from $\phi_{i}^b = \phi_{\text{min}}$ and $\tilde{f}_N=F$, the exponent term $g_{i,N}(m_{\rm h})$ for the last subcarrier is given by 
    %From \eqref{eq:exponent_term_stage_i_first}, it can be derived that the value of $\phi^{\rm b}_i$ that enforces $g_{i,h}(1) = 0$ is given by $\varphi_1$. In a similar manner, substituting $\varphi_1$ for $\phi^{\rm b}_i$ according to the result of \eqref{eq:exponent_term_stage_i_first} and recognizing that $\tilde{f}_N$ corresponds to $F$, $g_{i,h}(N)$ can be expressed as
    \begin{align}\label{eq:exponent_term_stage_i_last}
         g_{i,N}(m_{\rm h}) = &\big(m_{\rm h}-1 \big)\sin \hat{\theta}_i \bigg (\cos \varphi_N \Big ( 1+\frac{F}{f_c} \Big ) -\cos \phi_{\text{min}} \bigg ) \nonumber \\ &+ 2 F t^{\rm s,h}_{i,(m_{\rm h})}.
    \end{align}
    To ensure that the horizontal array gain of the last subcarrier is maximized at the maximum azimuth angle $\phi_{\text{max}}$, we set $\varphi_N = \phi_{\text{max}}$ and determine $t^{\rm s,h}_{i,(m_{\rm h})}$ as 
    \begin{align}\label{eq:TTD_mh_stage_i}
         t^{\rm s,h}_{i,(m_{\rm h})} = \frac{(m_{\rm h} - 1)\sin \hat{\theta}_i (\cos \phi_{\text{min}}- \cos\phi_{\text{max}} (1+ \frac{F}{f_c}))}{2F},
    \end{align}
    for all  $m_{\rm h} \in\{1,\ldots,M_{\rm h}\}$.
    Applying the values of $\phi^{\rm b}_i$ and $t^{\rm s,h}_{i,(m_{\rm h})}$ to \eqref{eq:exponent_term_mh_stage_i} and solving for $\varphi_n$ that ensures $g_{i,n}(m_{\rm h})=0$, the azimuth angle of subcarrier $n$ due to the beam-squint effect is determined as
    \begin{align}\label{eq:varphi_n}
        \varphi_{n} {=} \arccos \!\left ( \frac { \cos  \phi_{\text{min}} - \frac{\tilde{f}_n}{F} \Big (\cos \phi_{\text{min}} -  \cos\phi_{\text{max}} \big ({1 + \frac {F}{f_{c}}}\big ) \Big ) }{1+\frac {\tilde{f}_n}{f_{c}} } \right) \!.
    \end{align}
    As can be seen in \eqref{eq:varphi_n}, our sensing beamforming vector ${\bf b}_{i,n}$ in stage $i$ concurrently covers $N$ different azimuth angles in the ROI with the elevation angle $\hat{\theta}_i$, given by $\varphi_{1},\ldots, \varphi_{N}$, using $N$ subcarriers. Note that the azimuth grid structure remains unchanged regardless of the fixed elevation angle $\hat{\theta}$. This sensing approach allows us to reduce the scanning time to search for the azimuth angles of the $Q_i$ targets at the elevation angle $\hat{\theta}_i$. Details of the target detection strategy for stage $i$ will be described in Sec.~\ref{Sec:Detection}.% which will be clarified in Sec.~\ref{Sec:Detection}.  

    \subsection{Beamforming for Communication Users}\label{Sec:Comm_BF}  

    While the sensing beamforming exploits the beam-squint effect to cover the ROI, the communication beamforming is designed to achieve precise beam steering toward each CU location $(\theta_k^{\rm cu}, \phi_k^{\rm cu})$ across all subcarriers. The design follows a similar approach as in Sec.~\ref{Sec:Sensing_BF_AAS}, with the key distinction that the PS parameters do not need to be found, while the TTD parameters are optimized to maximize the array gain. The communication beamforming vector for the $k$-th CU on subcarrier $n$ is expressed as 
    \begin{align}\label{eq:Comm_BF}
        \mathbf w_{k,n} =  {\sf diag}( {\bf t}^{\rm c}_{k,n} ) \mathbf a^{\sf H}(\theta_k^{\rm cu}, \phi_k^{\rm cu},0),
    \end{align}
    where ${\bf t}^{\rm c}_{k,n}=\big [e^{-j2\pi \tilde{f}_n t^{\rm{c}}_{k, (1,1)} }, \cdots, e^{-j2\pi\tilde{f}_n t^{\rm{c}}_{k, (M_{\rm h},M_{\rm v})} } \big]$ \cite{beam_squint_split}.
    The array gain expression can be factorized into horizontal and vertical components, similar to the sensing beamforming design. To facilitate this factorization, we decompose the TTD parameter as $t_{k,(m_{\rm h},m_{\rm v})}^{\rm c}=t^{\rm c,h}_{k,(m_{\rm h})} + t^{\rm c,v}_{k,(m_{\rm v})}$ with ${\bf t}_{k,n}^{\rm c,h} = \big [e^{-j2\pi\tilde{f}_n t^{\rm c,h}_{k,(1)} }, \cdots, e^{-j2\pi\tilde{f}_n t^{\rm c,h}_{k,(M_{\rm h})} } \big]$ and ${\bf t}_{k,n}^{\rm c,v} = \big [e^{-j2\pi\tilde{f}_n t^{\rm c,v}_{k,(1)} }, \cdots, e^{-j2\pi\tilde{f}_n t^{\rm c,h}_{k,(M_{\rm v})} } \big]$.
    The optimal TTDs are determined by ensuring that the exponent terms in the array gain's summation form are set to zero. Since this derivation is similar to the one for the AAS process in Sec.~\ref{Sec:Sensing_BF_AAS}, the detailed mathematical steps are omitted for brevity. This procedure yields the following closed-form expressions:
    \begin{align}\label{eq:TTD_mh_comm}
         t^{\rm c,h}_{k,(m_{\rm h})} &= \frac{(1 - m_{\rm h})\sin\theta_k^{\rm cu} \cos \phi_k^{\rm cu}}{2f_c},~\forall m_{\rm h} \in\{1,\ldots,M_{\rm h}\}, \nonumber \\
         t^{\rm c,v}_{k,(m_{\rm v})} &= \frac{(1 - m_{\rm v})\cos \theta_k^{\rm cu}}{2f_c},~\forall m_{\rm v} \in\{1,\ldots,M_{\rm v}\}.
    \end{align}
    With these optimal parameter settings, the beamforming vector $\mathbf w_{k,n}$ maximizes the array gain at the CU’s angular location and effectively mitigates the beam-squint effect, ensuring consistent communication performance across the entire bandwidth.
    % \textcolor{red}{
    % Referring to \eqref{eq:Received_comm_signal}, the signal-to-interference-plus-noise ratio (SINR) for $k$-th CU on subcarrier $n$ in the $t$-th OFDM symbol of stage $i$ can be expressed as shown in Eq. \eqref{eq:Comm_SINR}.
    % \begin{figure*}
    % \begin{align}\label{eq:Comm_SINR}
    % \text{SINR}_{i,k,n} (t) =\frac { \left | \mathbf{h}_{n} (\theta_k, \phi_k) \mathbf w_{i,k,n} \right |^{2} \xi_{i,k,n} (t)} {  \sum _{l=1,l \neq k}^{K} \left | \mathbf{h}_{n} (\theta_k, \phi_k) \mathbf w_{i,l,n} \right |^2 \xi_{i,l,n} (t)+ \left | \mathbf{h}_{n} (\theta_k, \phi_k) \mathbf b_{i,n} \right |^2 \zeta_{i,n}  + \sigma^{2}_v}
    % \end{align}
    % \hrulefill	
    % \end{figure*}
    % The Eq. \eqref{eq:Comm_SINR} considers the average power of the desired communication signal in the numerator, while the denominator accounts for the average power for interference from other CUs, the sensing signal, and additive noise. In contrast to the repetitive structure of the probing signal, the communication signal sends different values over time, thereby necessitating an efficient and low-complexity power allocation strategy for the communication power embedded in each SINR during the sensing spreading.
    % }

    \section{Target Detection for the \\ Proposed Hierarchical Sensing Framework}\label{Sec:Detection} 
    In this section, we develop a target detection strategy for the proposed hierarchical sensing framework, which enables efficient detection of the elevation and azimuth angles of multiple targets.
    %\subsection{Target Detection based on Compressive Sensing}\label{Sec:Target_detection} %전면적인 수정
    %In the proposed framework, the sensing beamforming vector for stage $0$ and stage $i\in\{1,\ldots,I\}$ has the maximum array gain in $N$ grid points in the elevation and azimuth angles, respectively. 
    %Motivated by this fact, we develop an on-grid angle detection algorithm based on a compressive sensing approach. 
    %The key idea of our approach is on the received echo signals to estimate the locations of the targets. These allow for the localization of targets by utilizing only received echo signals across all subcarriers, achieved only by forming an analog beam with a single RF chain for sensing. 

    \subsection{Target Detection Problem}\label{Sec:Problem_formulation} 
    We start by defining a target detection problem in the proposed sensing framework.
    After applying the matched filtering to the received echo signal in \eqref{eq:Received_sensing_signal}, the matched echo signal $\tilde{r}_{i,n} (t)$ is expressed as 
    \begin{align}\label{eq:matched_simple_received_echo}
        \tilde{r}_{i,n}(t) &= \frac{u_{i,n}^*(t)}{|u_{i,n}(t)|} r_{i,n}(t) \nonumber \\ &= \sqrt{p_{i,n}^{\rm s}}\mathbf b^{\sf H}_{i,n} \mathbf G_{n} \mathbf b_{i,n} + \frac{u_{i,n}^*(t)}{|u_{i,n}(t)|}{\bf b}_{i,n}^{\sf H} \mathbf{v}_{i,n}(t) \nonumber \\
        &= \sqrt{\frac{\kappa}{1+\kappa}} \sum_{q=1}^{Q} \sqrt{p_{i,n}^{\rm s}} \alpha_q | {\bf a}(\theta_q^{\rm ta}, \phi_q^{\rm ta},\tilde{f}_n) {\bf b}_{i,n} |^2 e^{-j \frac{4\pi }{\lambda}l_q^{\rm ta}} \nonumber \\
        &~~~~ + \sqrt{\frac{1}{1+\kappa}} \! \frac{1}{\sqrt{C}} \!\! \sum_{c=1}^{C} \!\! \sqrt{p_{i,n}^{\rm s}} \alpha^{\rm clu}_c \eta_c | {\bf a}(\theta_c^{\rm clu} \!, \phi_c^{\rm clu} \!,\tilde{f}_n) {\bf b}_{i,n} |^2 \nonumber \\
        &~~~~ + \frac{u_{i,n}^*(t)}{|u_{i,n}(t)|}{\bf b}_{i,n}^{\sf H} \mathbf{v}_{i,n}(t)\nonumber \\
        &\approx \sum_{q=1}^{Q} \underbrace{\sqrt{p_{i,n}^{\rm s}} \alpha_q | {\bf a}(\theta_q^{\rm ta}, \phi_q^{\rm ta},\tilde{f}_n) {\bf b}_{i,n} |^2}_{\triangleq c_{i,n}(\theta_q^{\rm ta}, \phi_q^{\rm ta})} \underbrace{e^{-j \frac{4\pi }{\lambda}l_q^{\rm ta}}}_{\triangleq e_{q}} + \tilde{v}_{i,n} (t),
    \end{align}
    where $\tilde{v}_{i,n} \! (t) \!\! = \!\! \sqrt{\!\frac{1}{1+\kappa}} \! \frac{1}{\sqrt{C}} \!\!\sum_{c=1}^{C} \!\! \sqrt{p_{i,n}^{\rm s}} \alpha^{\rm clu}_c \eta_c | {\bf a}(\theta_c^{\rm clu} \!, \phi_c^{\rm clu} \!,\! \tilde{f}_n \!) {\bf b}_{i,n} |^2 \\ + \frac{u_{i,n}^*(t)}{|u_{i,n}(t)|} {\bf b}_{i,n}^{\sf H}\mathbf{v}_{i,n}(t)$ is an effective noise, $c_{i,n}(\theta_q^{\rm ta}, \phi_q^{\rm ta})$ denotes the sensing beamforming gain, and $e_q$ denotes the complex phasor associated with the $q$‑th target \cite{Matched_filter}. The approximation in \eqref{eq:matched_simple_received_echo} holds when $\sqrt{\frac{\kappa}{1+\kappa}} \approx 1$, indicating a LoS-dominant scenario. This condition is particularly suitable for our considered wideband scenario where beam squint effects are non-negligible.\footnote{Since the LoS path contains the target's angle information via the steering vector ${\bf a}(\theta^{\rm ta}, \phi^{\rm ta},\tilde{f})$, the LoS-dominant scenario is essential for formulating the angle-based target detection problem. This ensures angle-based target detection performance across the wideband system.}
    % Note that the effective channel for the $q$-th target can be decomposed into $c_{i,n}(\theta_q^{\rm ta}, \phi_q^{\rm ta})$, representing the beamforming gain without the phasor, and $e_q$, which specifically accounts for the phasor. Additionally, $\tilde{v}_{i,n}(t) \triangleq {\bf b}_{i,n}^{\sf H}\mathbf{v}_{i,n}(t) u_{i,n}^*(t)/|u_{i,n}(t)|$ is the effective noise signal.
    Given that the sensing signal is transmitted using $T_i$ OFDM symbols during stage $i$, an average echo signal is given by
    \begin{align}\label{eq:avg_simple_received_echo}
        \bar{r}_{i,n} =\frac{1}{T_i}\sum_{t=1}^{T_i}\tilde{r}_{i,n}(t) = \sum_{q=1}^{Q} c_{i,n}(\theta_q^{\rm ta}, \phi_q^{\rm ta}) e_q + \bar{v}_{i,n}, 
    \end{align}
    where $\bar{v}_{i,n}\triangleq\frac{1}{T_i}\sum_{t=1}^{T_i}\tilde{v}_{i,n}(t)$ is an average effective noise signal on subcarrier $n$ in stage $i$.
    By stacking the average echo signals across all subcarriers, an observation vector in stage $i$ is defined as
    \begin{align}\label{eq:stacking_echo_signal}
    \underbrace { \begin{bmatrix}
        \bar{r}_{i,1}  \\
        \bar{r}_{i,2}  \\
        \vdots \\
        \bar{r}_{i,N}  
    \end{bmatrix}}_{\triangleq\bar{\bf r}_{i}}&=\sum_{q=1}^{Q}
    \underbrace { \begin{bmatrix}
        c_{i,1}(\theta_q^{\rm ta}, \phi_q^{\rm ta}) \\
        c_{i,2}(\theta_q^{\rm ta}, \phi_q^{\rm ta}) \\
        \vdots \\
        c_{i,N}(\theta_q^{\rm ta}, \phi_q^{\rm ta})
    \end{bmatrix}}_{\triangleq{\bf c}_{i}(\theta_q^{\rm ta}, \phi_q^{\rm ta})}  e_q +
    \underbrace { \begin{bmatrix}
        \bar{v}_{i,1} \\
        \bar{v}_{i,2} \\
        \vdots \\
        \bar{v}_{i,N} 
    \end{bmatrix}}_{\triangleq\bar{\bf v}_{i}} \nonumber \\
    &=   \begin{bmatrix}
    | & ~ & | \\
        {\bf c}_{i}(\theta_1^{\rm ta}, \phi_1^{\rm ta}) & \cdots & {\bf c}_{i}(\theta_{Q}^{\rm ta}, \phi_{Q}^{\rm ta}) \\
        |&~& |
    \end{bmatrix}  \begin{bmatrix}
        e_1 \\
        \vdots \\
        e_{Q}
    \end{bmatrix}+ \bar{\bf v}_{i},
    \end{align}
    where ${\bf c}_{i}(\theta_q^{\rm ta}, \phi_q^{\rm ta})$ can be interpreted as a {\em sensing gain pattern} for the position $(\theta_q^{\rm ta}, \phi_q^{\rm ta})$ in stage $i$. 
    Therefore, our target detection problem is to estimate the ${Q}$ unknown target positions, $\{(\theta_q^{\rm ta}, \phi_q^{\rm ta})\}_q$, based on the observation vectors from $I+1$ stages, $\{\bar{\bf r}_i\}_{i}$. 
    % where $\tilde{\bf r}_{i}(t)$ is given by $\tilde{\bf r}_{i}(t) = \sum_{q=1}^Q \tilde{\bf h}_{i}(\theta_q^{\rm s}, \phi_q^{\rm s}) + \tilde{\bf v}_{i} (t)$. Stage $i$ average of $\tilde{\bf r}_{i}(t)$ over $T_i$ repetitions can be obtained as
    % \begin{align}\label{eq:avg_Matched_filter_echo_signal}
    %     \tilde{\bf r}_{i} &=\frac{1}{T_i}\sum_{t=1}^{T_i}\tilde{\bf r}_{i}(t) = \sum_{q=1}^Q \tilde{\bf h}_{i}(\theta_q^{\rm s}, \phi_q^{\rm s}) + \tilde{\bf v}_{i}.
    % \end{align}
    % where averaged noise is defined as $\tilde{\bf v}_{i} = \sum^{T_i}_{t=1} \tilde{\bf v}_{i} (t) / {T_i}$.

    \subsection{A Compressed Sensing Viewpoint}\label{Sec:CS_viewpoint}
    The target detection problem of the proposed sensing framework can be interpreted as a sparse signal recovery problem in compressed sensing. To demonstrate this, we consider a set of $L$ angle candidates, $\{(\check{\theta}_{l},\check{\phi}_{l})\}_{l=1}^L$, and construct a measurement matrix ${\bf C}_i = [{\bf c}_i(\check{\theta}_{1},\check{\phi}_{1}),\cdots,{\bf c}_i(\check{\theta}_{L},\check{\phi}_{L})]$ whose $l$-th column corresponds to the sensing gain pattern of the $l$-th angle candidate $(\check{\theta}_{l},\check{\phi}_{l})$.
    % To demonstrate this fact, consider a measurement matrix ${\bf C}_i = [{\bf c}_i(\check{\theta}_{1},\check{\phi}_{1}),\cdots,{\bf c}_i(\check{\theta}_{L},\check{\phi}_{L})]$, which is constructed from sensing gain patterns corresponding to L angle candidates, $\{(\check{\theta}_{l},\check{\phi}_{l})\}_{l}$.
    How to set the angle candidates will be discussed in the sequel.

    In an ideal scenario where the $Q$ targets are positioned exactly on $Q$ distinct angle candidates, we have  ${\bf c}_{i}(\theta_q^{\rm ta}, \phi_q^{\rm ta}) = {\bf c}_{i}(\check{\theta}_{l_q},\check{\phi}_{l_q})$, where $(\check{\theta}_{l_q},\check{\phi}_{l_q})$ is the angle candidate corresponding to the $q$-th target's position. Under this condition, the observation vector $\bar{\bf r}_i$ in \eqref{eq:stacking_echo_signal} can be expressed as
    \begin{align}\label{eq:CS_ideal}
        \bar{\bf r}_i &=\begin{bmatrix}
        | & ~ & | \\
        {\bf c}_{i}(\check{\theta}_{1},\check{\phi}_{1}) & \cdots & {\bf c}_{i}(\check{\theta}_{L},\check{\phi}_{L}) \\
        |&~& |
    \end{bmatrix}  \begin{bmatrix}
        e_{i,1}\\
        \vdots \\
        e_{i,L}
    \end{bmatrix}+ \bar{\bf v}_{i} \nonumber \\
        &= {\bf C}_i  {\bf e}_i +  \bar{\bf v}_{i},
    \end{align}
    where $e_{i,l}$ is the $l$-th entry of ${\bf e}_i$, and ${\bf e}_i \in\mathbb{C}^{L}$ is a sparse vector that has only $Q$ nonzero entries. To be more specific, the $l$-th entry of ${\bf e}_i$ is given by 
    \begin{align}\label{eq:element_e}
        e_{i,l} = \begin{cases}
            e_{q}, & \text{if}~l = l_q, \\
            0, & \text{otherwise}.
        \end{cases}
    \end{align}
    The column vector $[\mathbf C_i]_{:,l_q}$ of $\mathbf C_i$, corresponding to the index of the $q$-th nonzero term in vector $\mathbf e_i$, is ${\bf c}_{i}(\check{\theta}_{l_q},\check{\phi}_{l_q})$. The expression in \eqref{eq:CS_ideal} clearly shows that the positions of the $Q$ targets can be determined by recovering the $Q$-sparse vector ${\bf e}_i$ from its noisy observation $\bar{\bf r}_i$, a well-known sparse signal recovery problem in compressed sensing \cite{CS_book}. 

    In practice, the observation vector $\bar{\bf r}_i$ in \eqref{eq:stacking_echo_signal} can only be approximately expressed as
    \begin{align}\label{eq:CS_problem}
        \bar{\bf r}_i \approx {\bf C}_i  {\bf e}_i +  \bar{\bf v}_{i}.
    \end{align}
    This approximation arises because the $Q$ targets may not be exactly positioned on the predefined angle candidates.     
    Recall that the sensing beamforming vector $\mathbf b_{0,n}$ on subcarrier $n$ achieves its maximum gain at $(\vartheta_n, \phi)$, where $\vartheta_n$ is given in \eqref{eq:vartheta_n} for any $\phi \in [\phi_{\rm min}, \phi_{\rm max}]$. Based on this principle, the measurement matrix for stage $0$ is defined using the dense dictionary of $L$ elevation angle candidates as
    \begin{align}\label{eq:stage_0_measure_mtx}
        {\bf C}_0 = [{\bf c}_0(\vartheta_1,\phi),\cdots,{\bf c}_0(\vartheta_L,\phi)] \in \mathbb{R}^{N\times L}.
    \end{align}
    After estimating elevation angles in stage $0$, the measurement matrix for stage $i \in \{1,\ldots,I \}$ is also formed from dense dictionary of $L$ azimuth angle candidates for the estimated elevation $\hat{\theta}_i$ as
    \begin{align}\label{eq:stage_i_measure_mtx}
        {\bf C}_i = [{\bf c}_i(\hat{\theta}_i,\varphi_1),\cdots,{\bf c}_i(\hat{\theta}_i,\varphi_L)] \in \mathbb{R}^{N\times L},
    \end{align}
    where $\varphi_n$ is determined according to \eqref{eq:varphi_n}.

    While various compressed sensing techniques can be employed to recover the sparse vector $\mathbf e_i$, we adopt a modified MP algorithm normalizing the predicted coefficients for target detection \cite{MP}. The modified version operates as a greedy iterative method in the same manner as the conventional MP, and also includes an additional step of normalizing the entries of $\mathbf e_i$ because the predicted value represents the channel complex phasor. This choice is motivated by its low computational complexity, which is $\mathcal{O}(2 Q L N)$ for the overall process, and the ability to explicitly control the number of iterations using the known number of targets\footnote{By enforcing correlation-based or residual-energy stopping condition in MP, spurious targets can be rejected. This is particularly beneficial in the EAS process, where such criteria suppress error propagation and enhance robustness.}. To implement this, we introduce a counting vector, with the same length as $\mathbf e_i$, to store the indices corresponding to the highest correlations. The non-zero entries in the final counting vector, which serves as the output of the target detection algorithm, indicate that a target is detected at the corresponding grid point in stage $i$.

%Although the native angular resolution is determined by the number of actual subcarriers, $N$, the off-grid problem can be addressed.  the measurement matrix $\mathbf C_i$ from a dense dictionary of $L$ angle candidates, where $L \gg N$. This process generates a finer-grained, large-scale dictionary, which corresponds to a longer sparse vector $\mathbf e_i$. By providing a more densely populated set of potential angles, this dictionary minimizes the off-grid error, allowing for a more accurate sparse representation of the true target locations and ultimately leading to estimation performance that exceeds the fundamental resolution determined by the $N$ subcarriers. This minimization of the off-grid error is particularly critical in our high-frequency scenario. Due to the small wavelength, the phase of entries in $\mathbf e_i$ is highly sensitive to distance errors. Employing a large number of angle candidates, $L$, which results in a denser dictionary of candidate angles, is thus an effective strategy to reduce this distance error, which in turn stabilizes the phase information and significantly enhances performance.
%It is natural to set the number of the angle candidates, $L$, as the number of actual subcarriers, $N$. Nevertheless, 
    {\bf Remark (The effect of the number of angle candidates):}
    The resolution of the proposed target detection approach is controlled by the number of angle candidates, $L$, in the measurement matrix $\mathbf C_i$. By employing a more densely populated set of potential angles, the off-grid error can be effectively reduced, which enables a more accurate sparse representation of the true target locations and allows the estimation performance to exceed the fundamental resolution limit imposed by the $N$ subcarriers. This is particularly critical at high-frequency bands, where the small wavelength makes the phase of the entries in $\mathbf e_i$ extremely sensitive to even minor distance errors. Increasing $L$ thereby constructing a denser dictionary serves as an effective means to mitigate this sensitivity and achieve substantial performance gains. However, since the computational complexity of the modified MP algorithm is $\mathcal{O}(2QLN)$, the selection of $L$ should balance the trade-off between angular resolution and computational burden.

    \section{Power Allocation for the Proposed Hierarchical Sensing Framework}\label{Sec:Power} 
    In this section, we present a power allocation strategy for the proposed hierarchical sensing framework to ensure consistent performance for both sensing and communication.

    \subsection{Sensing Power Allocation}\label{Sec:Sensing_Power}
    To establish a power allocation strategy for sensing, we first analyze the SNR of the sensing grid when employing our sensing beamforming vector introduced in Sec.~\ref{Sec:Hierarchical}. The analysis begins by considering a single target located at an arbitrary grid point $(\vartheta, \varphi)$ and assuming only a LoS path. The corresponding received echo signal $z_{i,n} (t)$ on subcarrier $n$ at the $t$-th OFDM symbol in stage $i$ is then given by
    \begin{align}\label{eq:Received_sensing_signal_stage0}
        z_{i,n} (t) = \mathbf b^{\sf H}_{i,n} {\mathbf G^{\rm los}_{n} (\vartheta, \varphi)} \mathbf b_{i,n}u_{i,n}(t) + \mathbf{b}_{i,n}^{\sf H} \mathbf v_{i,n} (t).
    \end{align}
    Since the BS is aware of the transmitted sensing signal, the matched filter is applied to $z_{i,n} (t)$ to maximize the SNR, as in \cite{Matched_filter}. Then the matched echo signal is given by 
    \begin{align}\label{eq:Matched_filter_signal_stage0}
        \tilde{z}_{i,n} (t) \!=\! \frac{u_{i,n}^*(t)}{|u_{i,n}(t)|} z_{i,n}(t) \!=\! \sqrt{p_{i,n}^{\rm s}}\mathbf b^{\sf H}_{i,n} \mathbf G^{\rm los}_{n} (\vartheta, \varphi) \mathbf b_{i,n} \!+\! \tilde{v}_{i,n} (t).
    \end{align}
    % where $\tilde{v}_{i,n} (t) = {\bf b}_{i,n}^{\sf H}\mathbf{v}_{i,n}(t) \frac{u_{i,n}^*(t)}{|u_{i,n}(t)|}$ is an effective noise. 
    where $\tilde{v}_{i,n} (t)$ is an effective noise. Considering the fact that the sensing signal is transmitted using $T_i$ OFDM symbols during stage $i$, an averaged matched echo signal is derived as
    \begin{align}\label{eq:avg_Matched_filter_signal_stage0}
        \bar{z}_{i,n} = \frac{1}{T_i} \sum_{t=1}^{T_i}\tilde{z}_{i,n}(t) =\! \sqrt{p_{i,n}^{\rm s}}\mathbf b^{\sf H}_{i,n} \mathbf G^{\rm los}_{n} (\vartheta, \varphi) \mathbf b_{i,n} + \bar{v}_{i,n},
    \end{align}
   where $\bar{v}_{i,n}$ is an average effective noise signal. Due to the properties of the AWGN, the distribution of $\bar{v}_{i,n}$ is identical to that of a single component of $\mathbf{v}_{i,n}(t)$. The SNR for $\bar{z}_{i,n}$, denoted as ${\sf SNR}_{i,n}^{\rm s}(\vartheta, \varphi)$, is given by 
    \begin{align}\label{eq:Sensing_SNR_stagei}
        {\sf SNR}_{i,n}^{\rm s}(\vartheta, \varphi) =\frac{T_i p_{i,n}^{\rm s} \left | \mathbf{b}_{i,n}^{\sf H} \mathbf G^{\rm los}_{n} (\vartheta, \varphi) \mathbf b_{i,n}  \right |^{2} }{ \sigma^2_{\rm BS}},
    \end{align}
    where the numerator represents the average power of the reflected sensing signal and the denominator represents the average power of the noise.

    Based on the sensing SNR analysis, we formulate a sensing power minimization problem under total power and sensing SNR constraints.
    Let $\tau^\text{s}>0$ be a sensing SNR threshold required for guaranteeing sufficient sensing performance for all grids. 
    %determine the sensing power level for each subcarrier in each stage. The objective of our sensing power allocation strategy is to ensure a uniform SNR across all grid points at each stage. To this end, we introduce a sensing SNR threshold, denoted by $\tau^\text{s}>0$, and allocate the power to maintain SNR levels above this threshold across all grids.
    Then, our sensing power minimization problem is formulated as 
    \begin{subequations}
    \begin{align}
        ({\bf P1})~\underset{\{p_{i,n}^{\rm s},T_i\}_i}{\min}~&  \sum_{n=1}^N\sum_{i=0}^{I} T_ip_{i,n}^{\rm s}, \\
        \text{s.t.}~~~&{\sf SNR}_{0,n}^{\rm s}(\vartheta_n,\phi) \geq \tau^{\rm s},~\forall n,~\phi \in [\phi_{\rm min},\phi_{\rm max}], \\
        &{\sf SNR}_{i,n}^{\rm s}(\hat{\theta_i},\varphi_n) \geq \tau^{\rm s},~\forall n,~i \in\{1,\ldots,I\},  \\
        & \sum_{n=1}^{N} p_{i,n}^{\rm s} \leq P_{\rm max}^{\rm s},~\forall i, 
        \\& p_{i,n}^{\rm s} \geq 0,~T_i \in \mathbb{N},~\forall i, 
    \end{align}
    \end{subequations}
    where $P_{\rm max}^{\rm s}$ is a total sensing power constraint for each OFDM symbol. 
    The first constraint in ${\bf P1}$ ensures that the sensing beamforming vector ${\bf b}_{0,n}$ on subcarrier $n$ in stage $0$ provides a sufficient sensing SNR for the grid point $(\vartheta_n, \phi)$ when $\phi$ is in the ROI.  
    Similarly, the second constraint in ${\bf P1}$ ensures that the sensing beamforming vector ${\bf b}_{i,n}$ on subcarrier $n$ in stage $i$ offers a sufficient sensing SNR for the grid $(\hat{\theta}_i, \varphi_n)$. 
%     Since sensing power is involved in determining the communication SINR, the sensing power is first allocated at each stage, followed by the allocation of communication power.
% This condition is expressed as
%     \begin{align}
%         \text{SNR}_{i,n}  &\geq \tau^\text{s}, ~ \forall i,n.\label{eq:SNR_threshold} 
%     \end{align}
%     Recall that in stage $0$, the sensing beamforming vector ${\bf b}_{0,n}$ on subcarrier $n$ has the maximum beamforming gain for the grid $(\vartheta_n, \phi)$ for all $\phi\in[\phi_{\rm min}, \phi_{\rm max}]$. 
%     The sensing power allocation for subcarrier $n$ considering the repetition parameter can be expressed as

    The sensing power minimization problem in ${\bf P1}$ can be easily solved by inspection. First, substituting the sensing SNR expression in \eqref{eq:Sensing_SNR_stagei} into the first and second constraints in ${\bf P1}$ yields
    \begin{align}\label{eq:power_stage_0}
        T_0 p_{0,n}^{\rm s} \geq \frac{\tau^\text{s} \sigma^2_{\rm BS}}{ | \mathbf{b}_{0,n}^{\sf H} \mathbf G^{\rm los}_{n} (\vartheta_n, \phi) \mathbf b_{0,n} |^{2}},
    \end{align}
    and 
    \begin{align}\label{eq:power_stage_i}
        T_ip_{i,n}^{\rm s} \geq \frac{\tau^\text{s} \sigma^2_{\rm BS}}{ | \mathbf{b}_{i,n}^{\sf H} \mathbf G^{\rm los}_{n} (\hat{\theta}_i, \varphi_n) \mathbf b_{i,n} |^{2}},
            % \begin{cases} 
            %     {\frac{\tau^\text{s} \sigma^2_{\rm BS}}{ \left | \mathbf{b}_{i,n}^{\sf H} \mathbf G_{n} (\vartheta_n, \phi) \mathbf b_{i,n} \right |^{2}}, ~ i=0,\forall n} \\
            %     {\frac{\tau^\text{s} \sigma^2_{\rm BS}}{ \left | \mathbf{b}_{i,n}^{\sf H} \mathbf G_{n} (\hat{\theta}_i, \varphi_n) \mathbf b_{i,n} \right |^{2}}. ~ i \in \{ 1, \ldots, I \}, \forall n} 
            % \end{cases}
    \end{align}
    respectively. Aggregating \eqref{eq:power_stage_0} and the third constraint in ${\bf P1}$ yields 
    \begin{align}\label{eq:power_stage_0_ineq}
        P_{\rm max}^{\rm s} 
        \geq \sum_{n=1}^{N} p_{0,n}^{\rm s} \geq \sum_{n=1}^{N} \frac{\tau^\text{s} \sigma^2_{\rm BS}}{ T_0 | \mathbf{b}_{0,n}^{\sf H} \mathbf G^{\rm los}_{n} (\vartheta_n, \phi) \mathbf b_{0,n} |^{2}}.
    \end{align}
    Therefore, the optimal value of $T_0$ to minimize the sensing power is determined as
    \begin{align}\label{eq:T_0}
        T_0 = \left\lceil \sum_{n=1}^{N} \frac{\tau^\text{s} \sigma^2_{\rm BS}}{ P_{\rm max}^{\rm s}  | \mathbf{b}_{0,n}^{\sf H} \mathbf G^{\rm los}_{n} (\vartheta_n, \phi) \mathbf b_{0,n} |^{2}}\right\rceil,
    \end{align}
    and the power on subcarrier $n$ for stage $0$ is determined as 
    \begin{align}\label{eq:power_stage_0_final}
        p_{0,n}^{\rm s} =\frac{\tau^\text{s} \sigma^2_{\rm BS}}{ T_0  | \mathbf{b}_{0,n}^{\sf H} \mathbf G^{\rm los}_{n} (\vartheta_n, \phi) \mathbf b_{0,n} |^{2}}.
    \end{align}
    Similarly, aggregating \eqref{eq:power_stage_i} and the third constraint in ${\bf P1}$ yields 
    \begin{align}\label{eq:power_stage_i_ineq}
        P_{\rm max}^{\rm s} 
        \geq \sum_{n=1}^{N} p_{i,n}^{\rm s} \geq \sum_{n=1}^{N} \frac{\tau^\text{s} \sigma^2_{\rm BS}}{ T_i | \mathbf{b}_{i,n}^{\sf H} \mathbf G^{\rm los}_{n} (\hat{\theta}_i, \varphi_n) \mathbf b_{i,n} |^{2}}.
    \end{align}
    Therefore, the optimal value of $T_i$ to minimize the sensing power is determined as
    \begin{align}\label{eq:T_i}
        T_i = \left\lceil \sum_{n=1}^{N} \frac{\tau^\text{s} \sigma^2_{\rm BS}}{ P_{\rm max}^{\rm s} | \mathbf{b}_{i,n}^{\sf H} \mathbf G^{\rm los}_{n} (\hat{\theta}_i, \varphi_n) \mathbf b_{i,n} |^{2}} \right\rceil,
    \end{align}
    and the power level for stage $i\in\{1,\ldots,I\}$ is determined as 
    \begin{align}\label{eq:power_stage_i_final}
        p_{i,n}^{\rm s} = \frac{\tau^\text{s} \sigma^2_{\rm BS}}{T_i | \mathbf{b}_{i,n}^{\sf H} \mathbf G^{\rm los}_{n} (\hat{\theta}_i, \varphi_n) \mathbf b_{i,n} |^{2}}.
    \end{align}

    \subsection{Communication Power Allocation}
    To establish a power allocation strategy for communications, we first analyze the SINR of each CU when employing the proposed sensing framework with the sensing power determined in Sec.~\ref{Sec:Sensing_Power}.
    From \eqref{eq:Received_comm_signal}, the SINR for $k$-th CU on subcarrier $n$ in stage $i$, denoted by ${\sf SINR}_{i,k,n}^{\rm c}$, is expressed as 
    \begin{align}\label{eq:Comm_SINR}
        {\sf SINR}_{i,k,n}^{\rm c} =\frac { \chi_{k,k,n} p_{i,k,n}^{\rm c}} {  \sum _{l=1,l \neq k}^{K} \chi_{k,l,n}  p_{i,l,n}^{\rm c} + \tilde{\sigma}_{i,k,n}^2},
    \end{align}
    where the constants $\chi_{k,l,n}$ and $\tilde{\sigma}_{i,k,n}^2$ are defined as
        \begin{align}\label{eq:chi_sigma}
            \chi_{k,l,n} &=  \left | \mathbf{h}_{n} (\theta_k^{\rm cu}, \phi_k^{\rm cu}) \mathbf w_{l,n} \right |^2, \\
            \tilde{\sigma}_{i,k,n}^2 &= \left | \mathbf{h}_{n} (\theta_k^{\rm cu}, \phi_k^{\rm cu}) \mathbf b_{i,n} \right |^2 p_{i,n}^{\rm s}  + \sigma^{2}_k,
        \end{align}
    respectively. In the above expression, $\chi_{k,l,n}$ denotes the received beamforming gain at the $k$-th CU for the communication beamforming vector ${\bf w}_{l,n}$, and $\tilde{\sigma}_{i,k,n}^2$ denotes the effective noise power, which includes interference from both the sensing signal and the AWGN. As can be seen in \eqref{eq:Comm_SINR}, our SINR analysis characterizes the impact of the sensing power $ p_{i,n}^{\rm s}$ on communication performance for each subcarrier in every stage.
    % \begin{figure*}
    % \begin{align}\label{eq:Comm_SINR}
    %     {\sf SINR}_{i,k,n}^{\rm c} =\frac { \left | \mathbf{h}_{n} (\theta_k^{\rm cu}, \phi_k^{\rm cu}) \mathbf w_{k,n} \right |^{2} p_{i,k,n}^{\rm c}} {  \sum _{l=1,l \neq k}^{K} \left | \mathbf{h}_{n} (\theta_k, \phi_k) \mathbf w_{l,n} \right |^2 p_{i,l,n}^{\rm c} + \left | \mathbf{h}_{n} (\theta_k^{\rm cu}, \phi_k^{\rm cu}) \mathbf b_{i,n} \right |^2 p_{i,n}^{\rm s}  + \sigma^{2}_k}.
    % \end{align}
    % \hrulefill	
    % \end{figure*}    
    %The expression in \eqref{eq:Comm_SINR} shows that ${\sf SINR}_{i,k,n}^{\rm c}$ considers the average power of the desired communication signal in the numerator, while the denominator accounts for the interference power from signals intended for other CUs as well as the sensing signal. Therefore, 
    %In contrast to the repetitive structure of the probing signal, the communication signal sends different values over time, thereby necessitating an efficient and low-complexity power allocation strategy for the communication power embedded in each SINR during the sensing spreading.

    Based on the communication SINR analysis, we formulate a communication power minimization problem under communication SINR constraints.
    Let $\tau^\text{c}>0$ be the communication SINR threshold required to ensure sufficient communication performance for each CU \cite{digital2}. 
    Then, our communication power minimization problem on subcarrier $n$ in stage $i$ is devised as 
    \begin{subequations}\label{eq:comm_power_opt}
     \begin{align}
        ({\bf P2})~\underset{\{p_{i,k,n}^{\rm c}\}_k}{\min}~&  \sum_{k=1}^K p_{i,k,n}^{\rm c}, \\
        \text{s.t.}~~~&{\sf SINR}_{i,k,n}^{\rm c} \geq \tau^{\rm c},~\forall k,  \\
        & p_{i,k,n}^{\rm c} \geq 0,~\forall k. 
    \end{align}       
    \end{subequations}
    From \eqref{eq:Comm_SINR}, the problem ${\bf P2}$ can be reformulated as 
    \begin{subequations}\label{eq:comm_power_opt_simple}
    \begin{align}
        ({\bf P2}^\prime)~\underset{\{p_{i,k,n}^{\rm c}\}_k}{\min}~&  \sum_{k=1}^K p_{i,k,n}^{\rm c}, \\
        \text{s.t.}~~~&p_{i,k,n}^{\rm c} \geq \sum_{l\neq k} \frac{\tau^{\rm c}\chi_{k,l,n}}{\chi_{k,k,n} } p_{i,l,n}^{\rm c}  + \frac{\tau^{\rm c}\tilde{\sigma}_{i,k,n}^2}{\chi_{k,k,n} },~\forall k, \label{eq:comm_power_opt_simple_inequality}\\
        & p_{i,k,n}^{\rm c} \geq 0,~\forall k. \label{eq:comm_power_opt_simple_nonzero}
    \end{align}
    \end{subequations}
    The power can be minimized if the constraint in \eqref{eq:comm_power_opt_simple_inequality} holds with equality. This equality condition can be rewritten in a matrix form:
    %as given in \eqref{eq:LA_form}. 
    %\begin{figure*}
    \begin{align}
    \resizebox{8.6cm}{!}{%
      $\begin{aligned}\label{eq:LA_form}
        { \underbrace { \begin{bmatrix}
                1 & -\frac{\tau^{\text{c}} \chi_{1,2,n}}{\chi_{1,1,n}} & \cdots & -\frac{\tau^{\text{c}} \chi_{1,K,n}}{\chi_{1,1,n}} \\[4pt]
                -\frac{\tau^{\text{c}} \chi_{2,1,n}}{\chi_{2,2,n}} & 1 & \cdots & -\frac{\tau^{\text{c}} \chi_{2,K,n}}{\chi_{2,2,n}} \\[4pt]
                \vdots & \vdots & \ddots & \vdots \\[4pt]
                -\frac{\tau^{\text{c}} \chi_{K,1,n}}{\chi_{K,K,n}} & -\frac{\tau^{\text{c}} \chi_{K,2,n}}{\chi_{K,K,n}} & \cdots & 1
                \end{bmatrix} }_{\triangleq {\bf D}_n} \underbrace { \begin{bmatrix} p_{i,1,n}^{\rm c} \\[4pt] p_{i,2,n}^{\rm c} \\[4pt] \vdots \\[4pt] p_{i,K,n}^{\rm c} \end{bmatrix} }_{\triangleq {\bf p}_{i,n}} = \underbrace { \begin{bmatrix} \frac{\tau^{\text{c}} \tilde{\sigma}_{i,1,n}^2}{\chi_{1,1,n}} \\[4pt] \frac{\tau^{\text{c}} \tilde{\sigma}_{i,2,n}^2}{\chi_{2,2,n}} \\[4pt] \vdots \\[4pt] \frac{\tau^{\text{c}} \tilde{\sigma}_{i,K,n}^2}{\chi_{K,K,n}} \end{bmatrix}. }_{\triangleq {\bf s}_{i,n}}}
      \end{aligned}$
    }
    \end{align}
    In our scenario, CUs are sufficiently well separated, ensuring that the desired channel gain $\left | \mathbf{h}_{n} (\theta_k^{\rm cu}, \phi_k^{\rm cu}) \mathbf w_{k,n} \right |^2$ substantially dominates the sum of the interference powers $\left | \mathbf{h}_{n} (\theta_k^{\rm cu}, \phi_k^{\rm cu}) \mathbf w_{l,n} \right |^2$ whose terms are attenuated by beam-direction mismatch. This relationship ensures that the following diagonal-dominance condition is satisfied for each $k$-th CU:
    \begin{align}\label{eq:P_condition}
        \frac{1}{\tau^{\text{c}}}> \sum_{l \neq k} \frac{\chi_{k,l,n}}{\chi_{k,k,n}}=  \sum_{l \neq k} \frac{\left | \mathbf{h}_{n} (\theta_k^{\rm cu}, \phi_k^{\rm cu}) \mathbf w_{l,n} \right |^2}{\left | \mathbf{h}_{n} (\theta_k^{\rm cu}, \phi_k^{\rm cu}) \mathbf w_{k,n} \right |^2},~\forall k.
    \end{align}
    Under this condition, the Perron-Frobenius theorem \cite{perron_fro} guarantees that all components of the power vector $\mathbf p_{i,n}$ are positive, thereby satisfying the constraint \eqref{eq:comm_power_opt_simple_nonzero}.
    Furthermore, the Lévy–Desplanques theorem \cite{levy_des1} states that such a strictly diagonally dominant matrix is invertible. This theorem specifically guarantees that a matrix $\mathbf A$ is invertible if, for each row, the magnitude of the diagonal entry strictly exceeds the sum of the magnitudes of the off-diagonal entries. This condition is expressed as $|[{\mathbf A}]_{i,i}| > \sum_{j \neq i} |[{\mathbf A}]_{i,j}|$, $\forall i$. Since condition \eqref{eq:P_condition} ensures that $\mathbf D_n$ fulfills this strict diagonal dominance requirement, its invertibility is guaranteed. This invertibility allows the solution to the optimal power allocation problem to be simplified as follows:
    \begin{align}\label{eq:commun_power}
        {\bf p}_{i,n} = \mathbf{D}_n^{-1} {\bf s}_{i,n}.
    \end{align}
    Note that if the condition in \eqref{eq:P_condition} is not satisfied, it may indicate that the SINR constraint $\tau^{\rm c}$ is infeasible. In such cases, a practical approach is to reduce the SINR constraint $\tau^{\rm c}$ to a lower value until the condition in \eqref{eq:P_condition} holds. This adjustment enables the determination of the optimal communication power while still satisfying the revised, feasible SINR requirements.

    \section{Numerical Results}\label{Sec:Simulation}
    In this section, we evaluate the superiority of the proposed hierarchical sensing framework. The height of the BS is $40$ m, the carrier frequency is $f_c = 30$ GHz, and the transmission bandwidth is $F = 6$ GHz. The UPAs comprise $M_{\rm h}=64$ horizontal elements and $M_{\rm v}=64$ vertical elements.\footnote{While active TTDs offer higher resolution, passive TTDs can be employed as a power-efficient alternative. Although they may exhibit slightly degraded performance, passive TTDs provide the required delay while consuming negligible or zero direct current power.} Unless otherwise specified, the number of communication users, the number of subcarriers, and the fine-grid size are fixed to $K=2$, $N=128$, and $L=4096$, respectively. To evaluate the proposed framework under NLoS conditions, the Rician $K$-factor is set to $\kappa = 8$ dB for all simulations. The ROI is defined with respect to the BS, with an elevation angle ranging from $\theta_{\rm min} = 15^{\circ}$ to $\theta_{\rm max} = 70^{\circ}$ and an azimuth angle ranging from $\phi_{\rm min}= 30^{\circ}$ to $\phi_{\rm max} = 150^{\circ}$. The noise variance $\sigma^2_{\rm BS}$ and $\sigma^2_{k}$ are derived from the power spectral density of thermal noise, which is set to $-174$ dBm/Hz. The RCS of all targets is assumed to be $\sigma_{\text{RCS}} = 10$ dBsm.

    For a sensing performance metric, we consider an average distance error defined as
    \begin{align}\label{eq:avg_distance_error}
        \frac{1}{Q}\sum^Q_{q=1}{ \sqrt{(x_q {-} x^{\rm ta}_q)^2 + (y_q {-} y^{\rm ta}_q)^2}},
    \end{align}
    where $(x^{\rm ta}_q, y^{\rm ta}_q)$ denotes the ground truth location of the $q$-th target, given by $x^{\rm ta}_q = d^{\rm ta}_q \cos \phi_q^{\rm ta}$ and $y^{\rm ta}_q = d^{\rm ta}_q \sin\phi_q^{\rm ta}$, with $d^{\rm ta}_q = H \tan \theta_q^{\rm ta}$, and $(x_q, y_q)$ denotes the  estimated location, defined as $x_q = l_q \cos(\phi_q)$ and $y_q = l_q \sin(\phi_q)$, with $l_q = H \tan(\theta_q)$. For simplicity, we sort $\{(\hat{\theta}_{i},\hat{\phi}_{i,1}),\ldots,(\hat{\theta}_{i},\hat{\phi}_{i,Q_i})\}_{i=1}^{I}$ in ascending order, first by elevation angle and then by azimuth angle, to define pairs $(\theta_q, \phi_q)$. For the communication performance, we consider the average sum rate, defined as
    \begin{align}\label{eq:avg_sum_rate}
        \frac{1}{I+1} \sum^{I}_{i=0} \sum_k \sum_n \log(1+\text{SINR}^{\rm c}_{i,k,n}).
    \end{align}
    We quantify transmit power using the total sensing power and the average transmit power, given by $\sum^{I}_{i=0} T_i \sum_n p^{\rm s}_{i,n}$ and $\frac{1}{I+1}\sum^{I}_{i=0} \left( \sum_n p^{\rm s}_{i,n} + \sum_k \sum_n p^{\rm c}_{i,k,n} \right)T_i$. Finally, the energy efficiency (EE) is determined by the ratio of the average sum rate to the average transmit power.
    The sensing performance is evaluated by averaging instantaneous results over $10^3$ independent off-grid random
    target realizations within the ROI.
    
    % To evaluate the average performance, we conduct Monte Carlo (MC) simulations, wherein the targets' locations are randomly generated within the defined ROI for each MC iteration.

    % For performance comparison, we consider the following baseline methods.
    % \begin{itemize}
    
    %     \item {\bf Exhaustive:} This method adopts an exhaustive search over $N \times N$ grid points in the elevation–azimuth angle domain. To implement this, the TTD lines and PSs of the UPA are carefully configured to form a single high-gain beam per OFDM symbol, utilizing all subcarriers. As a result, this approach benefits from a substantial beamforming gain proportional to the number of subcarriers. However, it requires $N^2$ OFDM symbols to scan the entire grid, which results in a considerable scan time. Once all grid points have been scanned, the target locations are estimated by identifying those grid points with the highest sensing gains.
        
    %     \item {\bf Azimuth only:} This method adopts an exhaustive search over $N$ grid points in the azimuth angle domain, while leveraging the beam-squint effect to simultaneously cover $N$ grid points in the elevation angle domain. As such, it serves as a hybrid approach between the conventional exhaustive search and the proposed sensing framework. To implement this method, the TTD lines and PSs of the UPA are carefully configured to steer the beam toward a specific azimuth angle, while distributing the beams across $N$ distinct elevation angles. Consequently, this approach requires $N$ OFDM symbols to scan the entire azimuth angle grid. 
    % \end{itemize}
    
    \begin{figure}[t]
        \centering %\vspace{-3mm}
        \subfigure[$Q=1$]
        {\epsfig{file=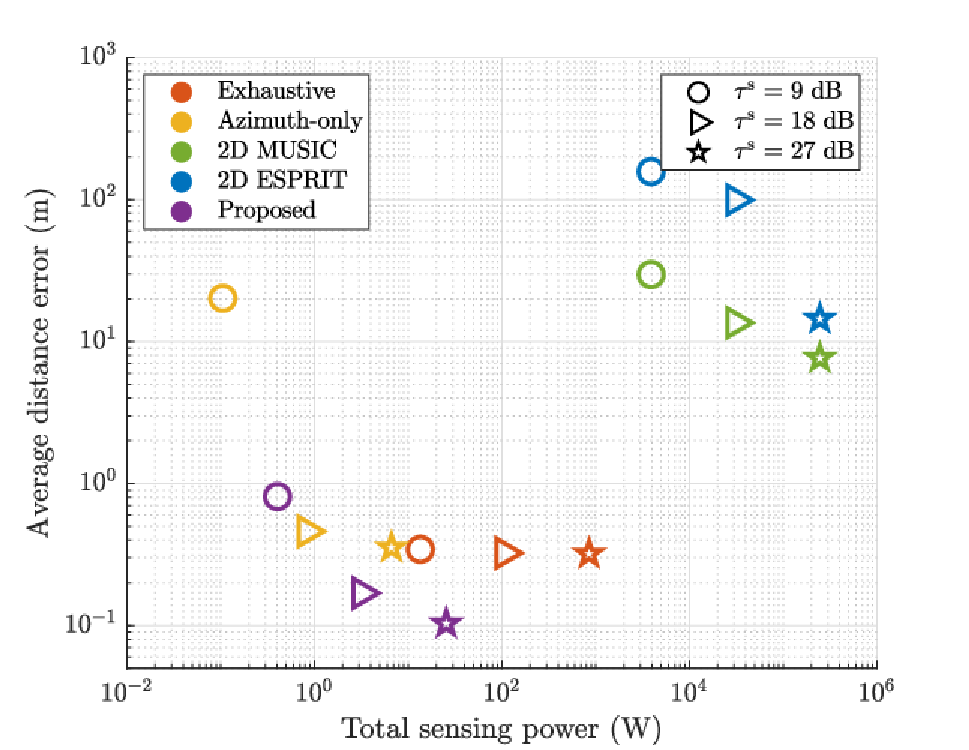, width=6.0cm}}
        \subfigure[$Q=3$]
        {\epsfig{file=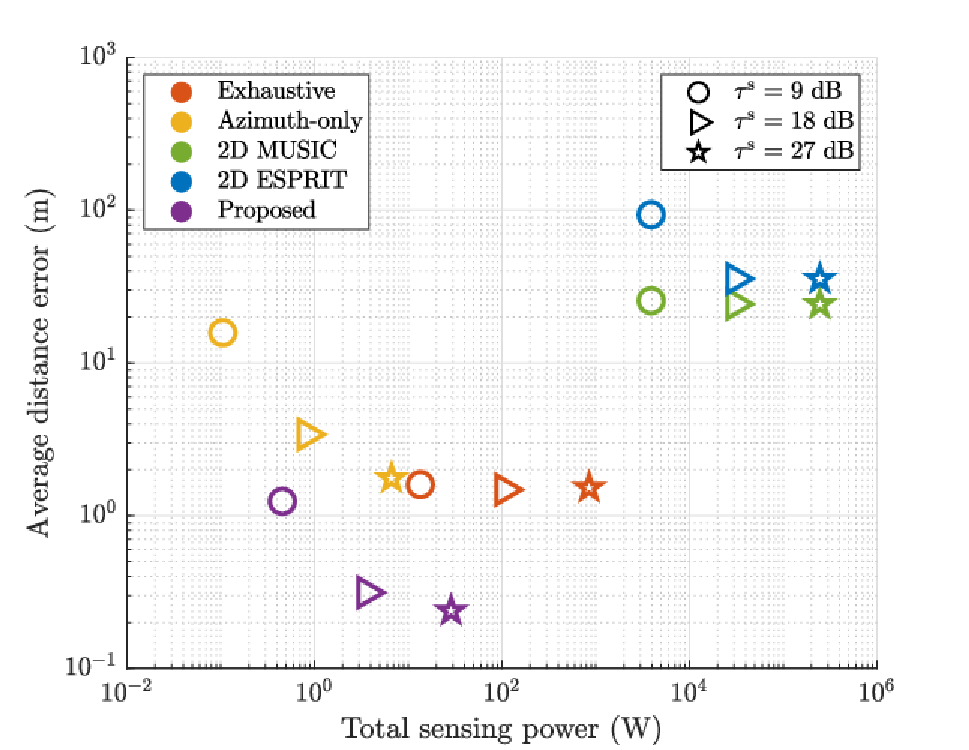, width=6.0cm}}
        \caption{The average distance error versus total sensing power of the proposed framework compared with subspace-based and exhaustive-based methods for $Q=1$ and $Q=3$.}\label{fig:comparison_existing}
        \vspace{-2mm}  %\vspace{-2mm}
    \end{figure}

    %\subsection{Performance Comparison} 
    %are extended for a 2D azimuth–elevation search, but they 
    Fig.~\ref{fig:comparison_existing} compares the sensing performance–transmit power trade-off of the proposed method against several baselines. In this simulation, we consider two subspace-based baselines, multiple signal classification (MUSIC) and estimation of signal parameters via rotational invariance techniques (ESPRIT), which are extended for a 2D elevation-azimuth search. We also consider two exhaustive search-based methods: (i) the Exhaustive method, which forms a single high-gain beam per OFDM symbol using all subcarriers to scan an $N \times N$ grid, thus requiring a total of $N^2$ OFDM symbols, and (ii) the Azimuth-only method, which scans $N$ grid points on the azimuth axis while exploiting the beam-squint effect to simultaneously cover $N$ grid points on the elevation axis, requiring only $N$ OFDM symbols in total and thus consuming less sensing power. The results from this comparison demonstrate that the proposed method achieves the most favorable performance–power trade-off among all baselines. In particular, both MUSIC and ESPRIT are unable to mitigate the severe beam-squint effect in the wideband scenario, leading to degraded performance. While the Exhaustive method attains competitive accuracy, it requires more than 30 times the transmit power of the proposed approach. Conversely, the Azimuth-only method consumes less power but suffers from substantially reduced accuracy. Overall, these results confirm the superiority of the proposed method in terms of both accuracy and power efficiency.
    % \textcolor{blue}{Fig.~\ref{fig:comparison_existing} compares the sensing performance of the proposed framework against several baseline methods. For subspace-based benchmarks, the multiple signal classification (MUSIC) and estimation of signal parameters via rotational invariance techniques (ESPRIT) algorithms were extended for a 2D search to estimate both azimuth and elevation angles. The results demonstrate that these methods exhibit poor sensing performance due to their inability to mitigate the severe beam-squint effect inherent in the considered wideband scenario. Moreover, they require substantially higher transmit power. In contrast, the exhaustive search-based methods, which are designed to account for the beam-squint effect, achieve better sensing accuracy than the subspace-based approaches, yet their performance remains inferior to our proposed framework. The Azimuth-only method consumes less transmit power than the Exhaustive method but at the cost of further degraded sensing accuracy. Crucially, the Exhaustive method requires over 30 times more transmit power than the proposed framework, underscoring the superior performance of the proposed framework in both accuracy and power efficiency.}
    
    \begin{figure}[t]
        \centering %\vspace{-3mm}
        {\epsfig{file=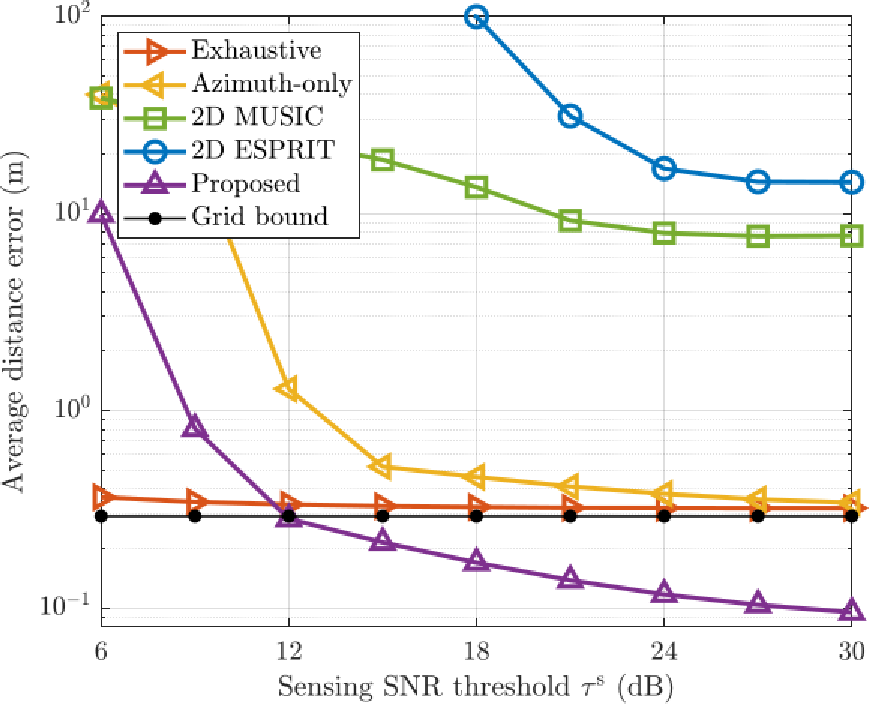, width=6.0cm}}
        \caption{The average distance error versus sensing SNR threshold $\tau^{\rm s}$ of the proposed framework compared with various sensing methods for $Q=1$.}\label{fig:comparison_conventional}
        \vspace{-2mm}
    \end{figure}

    Fig.~\ref{fig:comparison_conventional} illustrates the sensing performance of various sensing frameworks for different sensing SNR thresholds when $Q=1$. For reference, we also include the grid bound as a benchmark, defined as the average distance between the true target position and the nearest grid point. As discussed for Fig.~\ref{fig:comparison_existing}, the subspace-based baselines suffer from performance degradation as a result of the mismatch induced by frequency-dependent distortion, while the exhaustive search-based methods saturate at the grid bound. Notably, the results show that the proposed framework even outperforms the grid bound at high SNR thresholds, indicating that employing a large number of angle candidates (i.e., $L=4096$) enables the proposed method to achieve super-resolution performance.

    \begin{figure}[t]
        \centering %\vspace{-3mm}
        {\epsfig{file=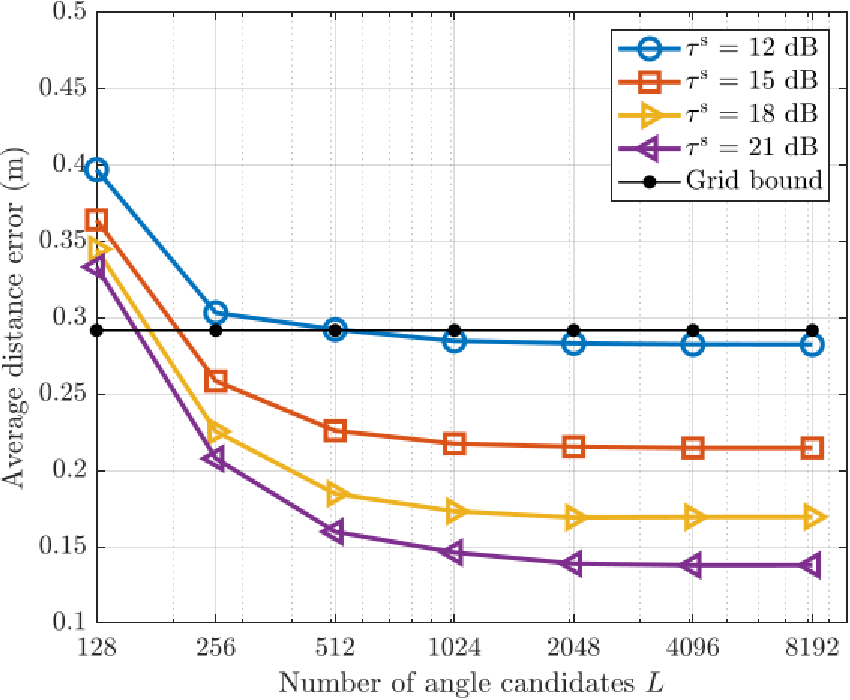, width=6.0cm}}
        \caption{Sensing performance versus the number of angle candidates $L$ under sensing SNR thresholds $\tau^{\rm s}$ from $12$ dB to $21$ dB for $Q=1$.}\label{fig:performance_vs_L}
        \vspace{-2mm}
    \end{figure}
    
    Fig.~\ref{fig:performance_vs_L} demonstrates the sensing performance for different numbers of angle candidates $L$. It is observed that the average distance error decreases as $L$ increases, whereas the performance gain becomes marginal beyond a certain value of $L$. Therefore, as mentioned in the {\bf Remark} of Sec.~\ref{Sec:CS_viewpoint}, $L$ should be selected to balance the trade-off between the computational complexity of the target detection algorithm and sensing performance. Furthermore, it is evident that the grid bound is surpassed when $L$ is set to a value larger than $L=N = 128$, and performance exceeding the grid bound can be achieved with smaller values of $L$ when the sensing SNR threshold $\tau^{\rm s}$ is high.

    \begin{figure}[t]
        \centering %\vspace{-3mm}
        {\epsfig{file=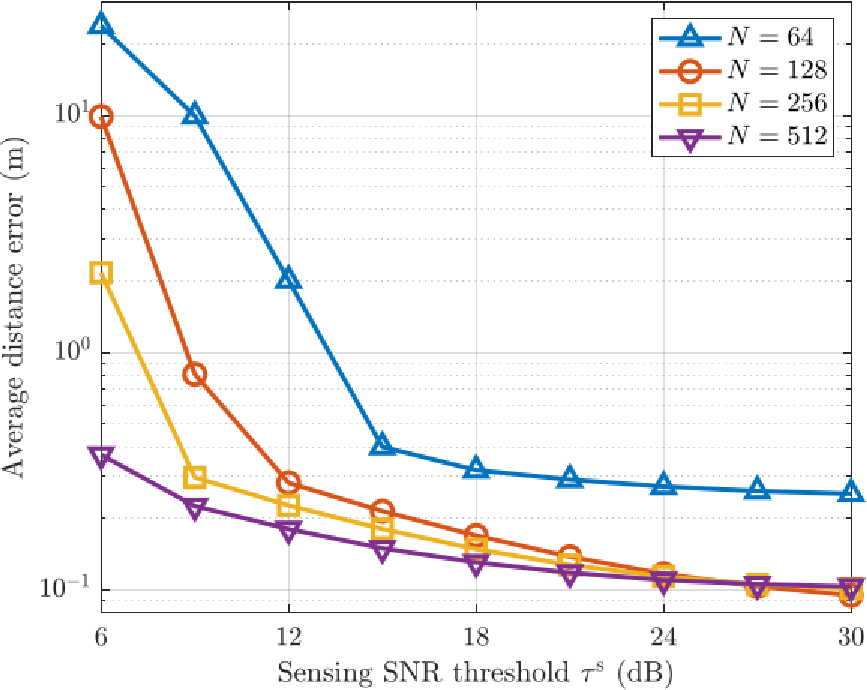,width=6.0cm}}%\vspace{-3mm}
        \caption{The average distance error versus sensing SNR threshold $\tau^{\rm s}$ achieved by the proposed framework with different numbers of subcarriers $N$ for $Q=1$.} \label{fig:num_subcarrier}
        \vspace{-2mm}
    \end{figure}
    
    Fig.~\ref{fig:num_subcarrier} compares the sensing performance of the proposed framework for different numbers of subcarriers when $Q=1$. The figure shows that as $N$ increases, the performance consistently improves due to the finer grid resolution from a larger number of subcarriers, and that the performance curves eventually converge to a similar point because all cases share the same number of angle candidates. On the other hand, the computational complexity of the target detection algorithm increases with $N$. Therefore, the performance gain from increasing $N$ is accompanied by an additional computational cost.

    \begin{figure}[t]
        \centering %\vspace{-3mm}
        {\epsfig{file=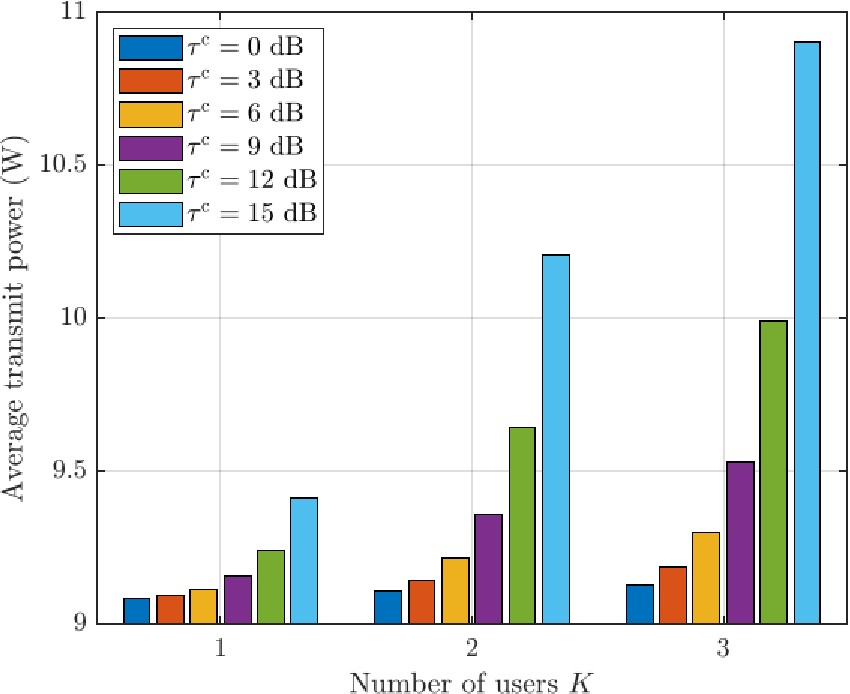,width=6.0cm}}%\vspace{-3mm}
        \caption{The average transmit power versus the number of communication users achieved by the proposed framework for $\tau^{\rm s} = 25~{\rm dB}$ and $Q=1$.}\label{fig:avg_transmit_power}
        % \caption{\textcolor{blue}{Average transmit power versus the number of communication users, parameterized by $\tau^{\rm c}$ under the fixed conditions of $\tau^{\rm s} = 25~{\rm dB}$ and $Q=1$.}}\label{fig:avg_transmit_power}
        \vspace{-2mm}
    \end{figure}
    
    Fig.~\ref{fig:avg_transmit_power} illustrates the average transmit power of the proposed framework as a function of the number of communication users for different values of the communication SINR threshold when $\tau^{\rm s} = 25~{\rm dB}$ and $Q=1$. As depicted in the figure, the average transmit power monotonically increases with the number of users. This trend is expected, as serving more communication users requires a larger total power allocation for communication. For a fixed $K$, the average transmit power also increases as $\tau^{\rm c}$ becomes larger. This is because more power must be allocated to meet the higher communication performance requirements imposed by a stricter SINR threshold.

    %antenna elements
    \begin{figure}[t]
        \centering %\vspace{-3mm}
        {\epsfig{file=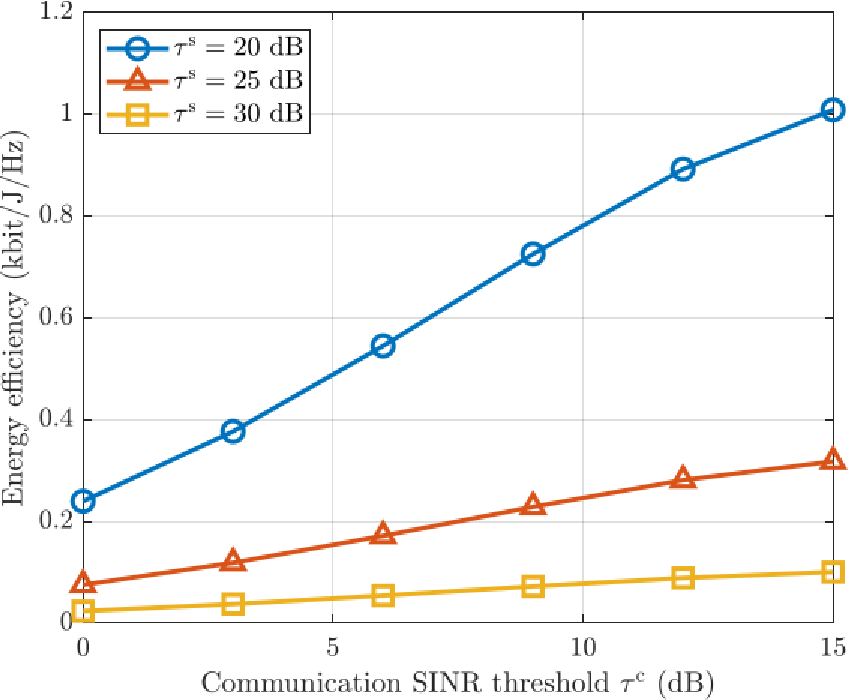,width=6.0cm}}%\vspace{-3mm}
        \caption{The energy efficiency versus communication SINR threshold achieved by the proposed framework for $Q=3$ and $K=3$.}\label{fig:EE}
        \vspace{-2mm}
    \end{figure}
    
    Fig.~\ref{fig:EE} illustrates the EE of the proposed framework as a function of the communication SINR threshold $\tau^{\rm c}$ when $Q=3$ and $K=3$. The results show a clear trend where the energy efficiency increases as $\tau^{\rm c}$ increases. This is because a higher $\tau^{\rm c}$ allows for a greater average sum rate, which is directly proportional to the EE. Conversely, an opposing trend is observed with respect to the sensing SNR threshold $\tau^{\rm s}$. As $\tau^{\rm s}$ increases, a higher total transmit power is required to meet the sensing quality constraint. This increased power consumption intuitively leads to a decrease in the overall energy efficiency.
    
    % trade-off
    \begin{figure}[t]
        \centering %\vspace{-3mm}
        {\epsfig{file=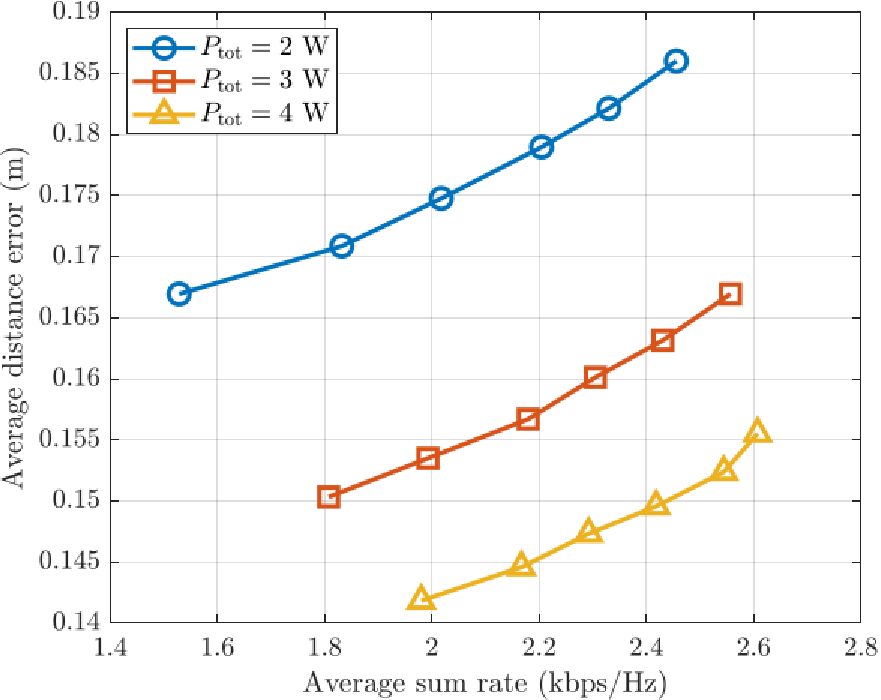,width=6.1cm}}%\vspace{-3mm}
        \caption{Trade-off between the average sum rate and the average distance error for $K=3$ and $Q=1$ under various average transmit power budgets $P_{\rm tot}$.} \label{fig:trade_off}
        \vspace{-2mm}
    \end{figure}
    
    % Fig.~\ref{fig:trade_off} illustrates the performance trade-off between sensing and communication achieved by the proposed framework for $K=3$ , $Q=1$, and $P_{\rm tot} =3$. In this simulation, each operating point is determined by varying the sensing SNR threshold and communication SINR threshold under a fixed total power budget of $P_{\rm tot}=3$W. The curve is upward-sloping because the y-axis represents the average distance error. This demonstrates that allocating more power to achieve a higher communication sum rate comes at the cost of increased sensing distance error. Conversely, allocating more power to meet a higher $\tau^{\rm s}$ reduces the distance error but directly lowers the achievable communication sum rate.

    Fig.~\ref{fig:trade_off} illustrates the performance trade-off between sensing performance and communication sum rate achieved by the proposed framework for $K=3$ and $Q=1$ across various average transmit power budgets $P_{\rm tot}$. Each operating point on the curves is derived by adjusting the sensing SNR threshold $\tau^{\rm s}$ and the communication SINR threshold $\tau^{\rm c}$ under a fixed $P_{\rm tot}$. The curve is upward-sloping because the y-axis represents the average distance error. This signifies the inherent trade-off in ISAC systems, where prioritizing a higher communication sum rate results in an increased sensing distance error, while increasing the sensing power to satisfy a higher $\tau^{\rm s}$ reduces the distance error but directly lowers the achievable communication sum rate. Notably, as $P_{\rm tot}$ increases, the available power for both functionalities expands, leading to a general improvement in the performance boundary. The results demonstrate that the system at $P_{\rm tot} = 4$ W simultaneously achieves lower sensing error and a higher communication sum rate compared to $P_{\rm tot} = 2$ W, proving that an increase in total transmit power effectively enhances the overall performance metrics of the system.
    
    % becomes dominant,  targets are randomly positioned off-grid when a target lies approximately the midpoint between grid points, a narrow beam may not capture it effectively, whereas a wide beam can cover a broader area around the grid point, thereby enhancing detection performance. 
    
    % A smaller UPA size results in a wider beam generated from the BS, whereas a larger size yields a narrower beam. For higher values of $\tau^{\rm s}$, performance closely approximates the grid bound, particularly when the UPA size is large. Conversely, at lower $\tau^{\rm s}$ values, performance improves with a smaller $M_{\rm v}$, corresponding to a wider beam. This behavior is attributed to the fact that targets are randomly positioned off-grid when a target lies approximately the midpoint between grid points, a narrow beam may not capture it effectively, whereas a wide beam can cover a broader area around the grid point, thereby enhancing detection performance. However, when $\tau^{\rm s}$ is sufficiently high, the signal strength is large enough that even a narrow beam can effectively cover off-grid targets.

    \section{Conclusion}\label{Sec:Conclusion}
    In this paper, we have proposed a hierarchical sensing framework for wideband ISAC with UPAs that exploits beam squint for efficient 2D angle estimation. By coordinating TTD lines and PSs, we have designed beamforming and grid structures to maximize array gain while enabling simultaneous multi grid point search. A compressed-sensing-based target detection problem has been formulated, and power-minimization policies for both sensing and communication have been developed to meet reliability targets. Simulation results demonstrate the superiority of the proposed framework over exhaustive-based and subspace-based approaches. Future work includes designing robust tracking that jointly estimates position and velocity for predictive beamforming, exploring cost-effective system architectures (e.g., sub-connected TTD), leveraging domain transform techniques for clutter suppression to enhance performance \cite{clutter_sup_1, clutter_sup_2}, and extending the framework to super-resolution recovery of off-grid targets via atomic norm minimization.

    \bibliographystyle{IEEEtran}
    \bibliography{Reference}

@ARTICLE{Interference,
  author={Tang, Aimin and Wang, Xudong and Zhang, J. Andrew},
  journal={IEEE Commun. Mag.}, 
  title={{Interference management for full-duplex ISAC in B5G/6G networks: Architectures, challenges, and solutions}}, 
  year={2024},
  volume={62},
  number={9},
  pages={20--26},
  month={Sep.}
}

@article{YOLO,
  title={{YOLO: An efficient terahertz band integrated sensing and communications scheme with beam squint}},
  author={Hongliang Luo and Feifei Gao and Hai Lin and Shaodan Ma and Poor, H Vincent},
  journal={IEEE Trans. Wireless Commun.},
  volume={23},
  number={8},
  pages={9389--9403},
  year={2024},
  month={Aug.}
}

@unpublished{levy_des1,
title={{AffineQuant: Affine transformation quantization for large language models}},
author={Yuexiao Ma and Huixia Li and Xiawu Zheng and Feng Ling and Xuefeng Xiao and Rui Wang and Shilei Wen and Fei Chao and Rongrong Ji},
year = {2024},
note = {arXiv:2403.12544}}

@ARTICLE{MP,
  author={Mallat, S.G. and Zhifeng Zhang},
  journal={IEEE Trans. Signal Process.}, 
  title={{Matching pursuits with time-frequency dictionaries}}, 
  year={1993},
  volume={41},
  number={12},
  pages={3397--3415},
  month={Dec.}
}

@ARTICLE{Matched_filter,
  author={Liu, Chunshan and Zhao, Lou and Li, Min and Yang, Lei},
  journal={IEEE Trans. Veh. Technol.}, 
  title={{Adaptive beam search for initial beam alignment in millimetre-wave communications}}, 
  year={2022},
  volume={71},
  number={6},
  pages={6801--6806},
  month={Jun.}
}

@ARTICLE{beam_squint_split,
  author={Gao, Feifei and Xu, Liangyuan and Ma, Shaodan},
  journal={IEEE Trans. Commun.}, 
  title={{Integrated sensing and communications with joint beam-squint and beam-split for mmWave/THz massive MIMO}}, 
  year={2023},
  volume={71},
  number={5},
  pages={2963--2976},
  month={May}
}

@ARTICLE{ideal_beam,
  author={Song, Jiho and Choi, Junil and Love, David J.},
  journal={IEEE Trans. Commun.}, 
  title={{Common codebook millimeter wave beam design: Designing beams for both sounding and communication with uniform planar arrays}}, 
  year={2017},
  volume={65},
  number={4},
  pages={1859--1872},
  month={Apr.}
}

@INPROCEEDINGS{beam_broad1,
  author={Fonteneau, Corentin and Crussière, Matthieu and Jahan, Bruno},
  booktitle={Proc. IEEE Joint Eur. Conf. Netw. Commun. 6G Summit}, 
  title={A systematic beam broadening method for large phased arrays}, 
  year={2021},
  pages={7--12},
  address={Porto, Portugal},
  month={Jun.}
}

@ARTICLE{beam_broad2,
  author={Xiao, Zhenyu and He, Tong and Xia, Pengfei and Xia, Xiang-Gen},
  journal={IEEE Trans. Wireless Commun.}, 
  title={{Hierarchical codebook design for beamforming training in millimeter-wave communication}}, 
  year={2016},
  volume={15},
  number={5},
  pages={3380--3392},
  month={May}
}

@ARTICLE{ISAC_appl1,
  author={Zhu, Xiaoqiang and Liu, Jiqiang and Lu, Lingyun and Zhang, Tao and Qiu, Tie and Wang, Chunpeng and Liu, Yuan},
  journal={IEEE Commun. Surveys Tuts.}, 
  title={{Enabling intelligent connectivity: A survey of secure ISAC in 6G networks}}, 
  year={2024},
  doi={10.1109/COMST.2024.3432871},
  month={early access, Jul. 24,}}

@ARTICLE{ISAC_appl2,
  author={Liu, Fan and Cui, Yuanhao and Masouros, Christos and Xu, Jie and Han, Tony Xiao and Eldar, Yonina C. and Buzzi, Stefano},
  journal={IEEE J. Sel. Areas Commun.}, 
  title={{Integrated sensing and communications: Toward dual-functional wireless networks for 6G and beyond}}, 
  year={2022},
  volume={40},
  number={6},
  pages={1728--1767},
  month={Jun.}}

@ARTICLE{IMT2030_1,
  author={Kaushik, Aryan and Singh, Rohit and Dayarathna, Shalanika and Senanayake, Rajitha and Di Renzo, Marco and Dajer, Miguel and Ji, Hyoungju and Kim, Younsun and Sciancalepore, Vincenzo and Zappone, Alessio and Shin, Wonjae},
  journal={IEEE Commun. Standards Mag.}, 
  title={{Toward integrated sensing and communications for 6G: Key enabling technologies, standardization, and challenges}}, 
  year={2024},
  volume={8},
  number={2},
  pages={52--59},
  month = {Jun.}}

@unpublished{IMT2030_2,
title={{Towards 6G evolution: Three enhancements, three innovations, and three major challenges}},
author={Rohit Singh and Aryan Kaushik and Wonjae Shin and Marco Di Renzo and Vincenzo Sciancalepore and Doohwan Lee and Hirofumi Sasaki and Arman Shojaeifard and Octavia A. Dobre},
year = {2024},
note = {arXiv:2402.10781}}

@ARTICLE{Multi_modal,
  author={Cheng, Xiang and Zhang, Haotian and Zhang, Jianan and Gao, Shijian and Li, Sijiang and Huang, Ziwei and Bai, Lu and Yang, Zonghui and Zheng, Xinhu and Yang, Liuqing},
  journal={IEEE Commun. Surveys Tuts.}, 
  title={{Intelligent multi-modal sensing-communication integration: Synesthesia of machines}}, 
  year={2024},
  volume={26},
  number={1},
  pages={258--301},
  month={1st Quart.}}

@ARTICLE{mutual_assist1,
  author={Meng, Kaitao and Masouros, Christos and Petropulu, Athina P. and Hanzo, Lajos},
  journal={IEEE Wireless Commun.}, 
  title={{Cooperative ISAC networks: Opportunities and challenges}}, 
  year={2024},
  pages={1--8},
  month={early access, Oct. 22, },
  doi={10.1109/MWC.008.2400151}}

@ARTICLE{mutual_assist2,
  author={Meng, Kaitao and Wu, Qingqing and Masouros, Christos and Chen, Wen and Li, Deshi},
  journal={IEEE Wireless Commun.}, 
  title={Intelligent surface empowered integrated sensing and communication: From coexistence to reciprocity}, 
  year={2024},
  volume={31},
  number={5},
  pages={84-91},
  month={Oct.}}

@ARTICLE{hybrid1,
  author={Cheng, Ziyang and He, Zishu and Liao, Bin},
  journal={IEEE J. Sel. Topics Signal Process.}, 
  title={{Hybrid beamforming design for OFDM dual-function radar-communication system}}, 
  year={2021},
  volume={15},
  number={6},
  pages={1455--1467},
  month={Nov.}}

@ARTICLE{digital1,
  author={Hua, Haocheng and Xu, Jie and Han, Tony Xiao},
  journal={IEEE Trans. Veh. Technol.}, 
  title={{Optimal transmit beamforming for integrated sensing and communication}}, 
  year={2023},
  volume={72},
  number={8},
  pages={10588--10603},
  month={Aug.}}

@ARTICLE{digital2,
  author={He, Zhenyao and Xu, Wei and Shen, Hong and Ng, Derrick Wing Kwan and Eldar, Yonina C. and You, Xiaohu},
  journal={IEEE J. Sel. Areas Commun.}, 
  title={{Full-duplex communication for ISAC: Joint beamforming and power optimization}}, 
  year={2023},
  volume={41},
  number={9},
  pages={2920--2936},
  month={Sep.}}

@ARTICLE{hybrid3,
  author={Wang, Xinyi and Fei, Zesong and Zhang, J. Andrew and Xu, Jie},
  journal={IEEE Trans. Commun.}, 
  title={{Partially-connected hybrid beamforming design for integrated sensing and communication systems}}, 
  year={2022},
  volume={70},
  number={10},
  pages={6648--6660},
  month={Oct.}}

@INPROCEEDINGS{ULA2,
  author={Liao, Bin and Ngo, Hien Quoc and Matthaiou, Michail and Smith, Peter J.},
  booktitle={Proc. IEEE Global Commun. Conf. (GLOBECOM)}, 
  title={{Low-complexity transmit beamforming design for massive MIMO-ISAC systems}}, 
  year={2023},
  pages={540--545},
  address={Kuala Lumpur, Malaysia},
  month={Dec.}}

@ARTICLE{ULA3,
  author={Zhang, Xiaoqi and Yuan, Weijie and Liu, Chang and Wu, Jun and Ng, Derrick Wing Kwan},
  journal={IEEE J. Sel. Topics Signal Process.}, 
  title={{Predictive beamforming for vehicles with complex behaviors in ISAC systems: A deep learning approach}}, 
  year={2024},
  volume={18},
  number={5},
  pages={828--841},
  month={Jul.}}

@ARTICLE{ULA5,
  author={Choi, Jinseok and Park, Jeonghun and Lee, Namyoon and Alkhateeb, Ahmed},
  journal={IEEE Trans. Wireless Commun.}, 
  title={{Joint and robust beamforming framework for integrated sensing and communication systems}}, 
  year={2024},
  volume={23},
  number={11},
  pages={17602--17618},
  month={Nov.}}

@ARTICLE{UPA1,
  author={Nguyen, Nhan Thanh and Nguyen, Van-Dinh and Nguyen, Hieu V. and Ngo, Hien Quoc and Swindlehurst, A. Lee and Juntti, Markku},
  journal={IEEE Trans. Signal Process.}, 
  title={{Performance analysis and power allocation for massive MIMO ISAC systems}}, 
  year={2025},
  pages={1--16},
  month={early access, Mar. 26,},
  doi={10.1109/TSP.2025.3554012}}

@ARTICLE{UPA2,
  author={Chu, Xinghe and Lu, Zhaoming and Kang, Jiawen and Zou, Yong and Zhang, Hui and Qiu, Xuesong},
  journal={IEEE Trans. Wireless Commun.}, 
  title={{Hybrid beamforming toward positioning enhancement under cellular MIMO systems}}, 
  year={2024},
  volume={23},
  number={10},
  pages={13545--13561},
  month={Oct.}}

@unpublished{UPA3,
title={{Beamforming design and multi-user scheduling in transmissive RIS enabled distributed cooperative ISAC networks with RSMA}},
author={Rohit Singh and Aryan Kaushik and Wonjae Shin and Marco Di Renzo and Vincenzo Sciancalepore and Doohwan Lee and Hirofumi Sasaki and Arman Shojaeifard and Octavia A. Dobre},
year = {2024},
note = {arXiv:2411.10960}}

@ARTICLE{wideband,
  author={Elbir, Ahmet M. and Mishra, Kumar Vijay and Chatzinotas, Symeon},
  journal={IEEE J. Sel. Topics Signal Process.}, 
  title={{Terahertz-band joint ultra-massive MIMO radar-communications: Model-based and model-free hybrid beamforming}}, 
  year={2021},
  volume={15},
  number={6},
  pages={1468--1483},
  month = {Nov.}}

@ARTICLE{beam_squint_solve,
  author={Elbir, Ahmet M. and Mishra, Kumar Vijay and Celik, Abdulkadir and Eltawil, Ahmed M.},
  journal={IEEE Commun. Mag.}, 
  title={{The curse of beam-squint in ISAC: Causes, implications, and mitigation strategies}}, 
  year={2024},
  volume={62},
  number={9},
  pages={52--58},
  month={Sep.}}

@ARTICLE{use_beam_squint1,
  author={Li, Jinyang and Zhang, Shun and Li, Zan and Ma, Jianpeng and Dobre, Octavia A.},
  journal={IEEE Trans. Commun.}, 
  title={{User sensing in RIS-aided wideband mmWave system with beam-squint and beam-split}}, 
  year={2025},
  volume={73},
  number={2},
  pages={1304--1319},
  month={Feb.}}

@INPROCEEDINGS{use_beam_squint2,
  author={Luo, Hongliang and Gao, Feifei and Yuan, Wanmai},
  booktitle={Proc. IEEE Int. Conf. Commun. (ICC)}, 
  title={{Near-field localization based on beam squint of mmWave communications}}, 
  year={2023},
  pages={6461--6466},
  month={May},
  addres={Rome, Italy}}

@ARTICLE{use_beam_squint3,
  author={Lei, Hao and Zhang, Jiayi and Xiao, Huahua and Ng, Derrick Wing Kwan and Ai, Bo},
  journal={IEEE Trans. Wireless Commun.}, 
  title={{Deep learning-based near-field user localization with beam squint in wideband XL-MIMO systems}}, 
  year={2025},
  volume={24},
  number={2},
  pages={1568--1583},
  month={Feb.}}

@ARTICLE{use_beam_squint4,
  author={Luo, Hongliang and Gao, Feifei and Yuan, Wanmai and Zhang, Shun},
  journal={IEEE Trans. Wireless Commun.}, 
  title={{Beam squint assisted user localization in near-field integrated sensing and communications systems}}, 
  year={2024},
  volume={23},
  number={5},
  pages={4504--4517},
  month={May}}

@INPROCEEDINGS{use_beam_squint_UPA2,
  author={Liu, Yilong and Zhang, Jun and Han, Yu and Wang, Jue and Jin, Shi and Li, Xiao},
  booktitle={Proc. 16th Int. Conf. Wireless Commun. Signal Process.}, 
  title={{Trade-off for terahertz beam squint enabled ISAC system}}, 
  year={2024},
  pages={294--298},
  month={Oct.},
  address={Hefei, China}}

@unpublished{use_beam_squint_UPA1,
title={{Synesthesia of machine (SoM)-driven analog precoder optimization for enhanced ISAC performance in sub-THz systems}},
author={Zonghui Yang and Shijian Gao and Xiang Cheng},
year = {2025},
note = {arXiv:2412.13532}}

@unpublished{user_loca1,
title={{Optimal beamforming for multi-target multi-user ISAC exploiting prior information: How many sensing beams are needed?}},
author={Jiayi Yao and Shuowen Zhang},
year = {2025},
note = {arXiv:2503.03560}}

@ARTICLE{user_loca2,
  author={Chen, Xu and Feng, Zhiyong and Zhang, J. Andrew and Wei, Zhiqing and Yuan, Xin and Zhang, Ping and Peng, Jinlin},
  journal={IEEE Trans. Veh. Technol.}, 
  title={{Downlink and uplink cooperative joint communication and sensing}}, 
  year={2024},
  volume={73},
  number={8},
  pages={11318--11332},
  month={Aug.}}

@book{CS_book,
  title={{Compressed sensing: Theory and applications}},
  author={Eldar, Yonina C and Kutyniok, Gitta},
  year={2012},
  publisher={\!\!\!\!\!\! Cambridge University Press}
}

@ARTICLE{drawback_digital,
  author={Molisch, Andreas F. and Ratnam, Vishnu V. and Han, Shengqian and Li, Zheda and Nguyen, Sinh Le Hong and Li, Linsheng and Haneda, Katsuyuki},
  journal={IEEE Commun. Mag.}, 
  title={{Hybrid beamforming for massive MIMO: A survey}}, 
  year={2017},
  volume={55},
  number={9},
  pages={134--141},
  month={Sep.}}

@ARTICLE{perron_fro,
  author={Pillai, S.U. and Suel, T. and Seunghun Cha},
  journal={IEEE Signal Process. Mag.}, 
  title={{The Perron-Frobenius theorem: Some of its applications}}, 
  year={2005},
  volume={22},
  number={2},
  pages={62--75},
  month={Mar.}}

@ARTICLE{clutter_sup_1,
  author={Yin, Jiapeng and Unal, Christine and Schleiss, Marc and Russchenberg, Herman},
  journal={IEEE Trans. Geosci. Remote Sens.}, 
  title={{Radar target and moving clutter separation based on the low-rank matrix optimization}}, 
  year={2018},
  volume={56},
  number={8},
  pages={4765--4780},
  month={Aug.}}

@ARTICLE{clutter_sup_2,
  author={Zhang, Shiyuan and Lu, Xingyu and Yu, Jintao and Dai, Zheng and Su, Weimin and Gu, Hong},
  journal={IEEE Geosci. Remote Sens. Lett.}, 
  title={{Clutter suppression for radar via deep joint sparse recovery network}}, 
  year={2024},
  volume={21},
  pages={1--5},
  month={Dec.}}

@INPROCEEDINGS{GC_Jo,
  author={Jo, Jaehong and Park, Jihun and Jeon, Yo-Seb and Poor, H Vincent},
  booktitle={IEEE Global Commun. Conf. (GLOBECOM)}, 
  title={{Beamforming design for hierarchical sensing in wideband integrated sensing and communications}}, 
  year={2025},
  address={Taipei, Tiwan},
  month={Dec.}}
    
\end{document}